\documentclass[
twocolumn,
amsmath,amssymb,
nofootinbib,
showkeys,
prd,
aps,
10pt,
longbibliography,
]{revtex4-2}

\usepackage{hyperref}
\usepackage{xcolor,graphicx}% Include figure files
\usepackage{bm}% bold math

\usepackage{soul}

\begin{document}
\interfootnotelinepenalty=10000

\title{From spacetime thermodynamics to Weyl transverse gravity}

\author{Ana Alonso-Serrano}
\email{ana.alonso.serrano@aei.mpg.de}
\affiliation{Institut für Physik, Humboldt-Universität zu Berlin, Zum Großen Windkanal 6, 12489 Berlin, Germany}
\affiliation{Max-Planck-Institut f\"ur Gravitationsphysik (Albert-Einstein-Institut), \\Am M\"{u}hlenberg 1, 14476 Potsdam, Germany}
	
\author{Luis J. Garay}
\email{luisj.garay@ucm.es}
\affiliation{Departamento de F\'{i}sica Te\'{o}rica and IPARCOS, Universidad Complutense de Madrid, 28040 Madrid, Spain}
	
\author{Marek Li\v{s}ka,}
\email{liskama4@stp.dias.ie}
\affiliation{School of Theoretical Physics, Dublin Institute for Advanced Studies, 10 Burlington Road, Dublin 4, Ireland}
\affiliation{Institute of Theoretical Physics, Faculty of Mathematics and Physics, Charles University, V Hole\v{s}ovi\v{c}k\'{a}ch 2, 180 00 Prague 8, Czech Republic}

\begin{abstract}
There exist two consistent theories of self-interacting gravitons: general relativity and Weyl transverse gravity. The latter has the same classical solutions as general relativity, but different local symmetries. We argue that Weyl transverse gravity also naturally arises from thermodynamic arguments. In particular, we show that thermodynamic equilibrium of local causal diamonds together with the strong equivalence principle encodes the gravitational dynamics of Weyl transverse gravity rather than general relativity. We obtain this result in a self-consistent way, verifying the validity of our initial assumptions, i.e. the proportionality between entropy and area and the different versions of the equivalence principle in Weyl transverse gravity. Furthermore, we extend the thermodynamic derivation of the equations of motion from Weyl transverse gravity to a class of modified theories of gravity with the same local symmetries. For this purpose, we employ the general expression for Wald entropy in such theories.
\end{abstract}
	
\keywords{Thermodynamics of spacetime, causal diamonds, entanglement entropy, Wald entropy, Weyl transverse gravity, unimodular gravity}

	\maketitle
	\tableofcontents

\section{Introduction}

Gravitational dynamics is connected to thermodynamics in a way that has not been observed for other physical theories~\cite{Bekenstein:1973,Hawking:1975,Braden:1990,Jacobson:1995ab,Jacobson:2003,Padmanabhan:2010,Jacobson:2015}. This connection becomes especially apparent in the entropy expression entering the laws of black hole thermodynamics~\cite{Bardeen:1973,Gibbons:1977,Wald:1993,Wald:1994}. The gravitational entropy associated with a Killing horizon of a black hole (as well as with other types of causal horizons~\cite{Gibbons:1977,Jacobson:2003,Jacobson:2019}) corresponds to a conserved Noether charge associated with the Killing symmetry. This charge is directly determined by the total divergence part of the variation of the gravitational Lagrangian~\cite{Wald:1993,Wald:1994,Compere:2018,Margalef:2021,Alonso:2022,Alonso:2022b,Hollands:2024}. Remarkably, the Noether charge not only determines the entropy, but also contains enough information to reconstruct the equations of motion of the gravitational theory~\cite{Padmanabhan:2008,Padmanabhan:2010,Jacobson:2012,Bueno:2017,Svesko:2018,Svesko:2019} (for purely metric theories whose Lagrangians do not contain the derivatives of the Riemann tensor). In other words, gravitational dynamics straightforwardly determines the expression for entropy of a horizon and this entropy in turn suffices to determine gravitational dynamics.

Our present paper is inspired by this strong relation between gravity and entropy and, in particular, by the seminal paper on the recovery of gravitational dynamics from thermodynamics~\cite{Jacobson:1995ab}. However, we take the correspondence between thermodynamics and gravity farther and \textit{assume} that thermodynamics of locally constructed causal horizons encodes \textit{all} the information about gravitational dynamics. We then show that this assumption, if taken seriously, leads to new insights into the nature of gravity.

To be more precise, we derive the equations governing the gravitational dynamics from the following two requirements:
\begin{itemize}
\item We assign to any causal horizon an entropy proportional to the area of its spatial cross-section. This form of entropy associated with a horizon is consistent with Bekenstein entropy formula valid for black holes in general relativity~\cite{Bekenstein:1973}. It also agrees with the behavior of vacuum entanglement entropy~\cite{Sorkin:1986,Srednicki:1993,Solodukhin:2011}. We reserve a more complete discussion of the naturalness of this assumption for appendix~\ref{lcd entropy}.
\item We impose the strong equivalence principle i.e., that all test fundamental physics (including gravitational physics) is locally unaffected by the presence of a gravitational field. This version of the equivalence principle allows us to derive the equations governing the gravitational dynamics locally and then extend the result to the entire spacetime.

\end{itemize}

A clarification is due for readers intimately familiar with the thermodynamics of spacetime program. The minimal requirement to recover gravitational dynamics from thermodynamics is actually the Einstein equivalence principle, which does not apply to self-gravitating bodies. However, the Einstein equivalence principle leaves room for the areal density of horizon entropy to depend on the position in the spacetime~\cite{Chirco:2010}. Then, we do not recover the (traceless) Einstein equations, but equations that contain some higher order corrections depending on the precise form of the areal density of entropy~\cite{Eling:2006,Jacobson:2012,Bueno:2017,Svesko:2018}. Modified theories of gravity are indeed not compatible with the strong equivalence principle~\cite{Casola:2014}. Therefore, we assume the strong equivalence principle to ensure the recovery of the lowest order gravitational dynamics, governed by the traceless Einstein equations. In section~\ref{WTDiff from TD}, we then relax to the Einstein equivalence principle in order to study the modified theories of gravity.

The technical implementation of these two assumptions is rather involved and we describe the necessary tools in the following sections. However, the key physical insights we arrive at in this work follow from the two points stated above and they are independent of the technical details.

Rather surprisingly, taking entropy proportional to area and invoking the strong equivalence principle does not lead us to general relativity, even though both features are characteristic of this theory. Instead, we recover a gravitational dynamics consistent with Weyl transverse gravity~\cite{Alvarez:2006,Carballo:2022}. This theory has the same classical solutions as general relativity, but its equations of motion are traceless and, rather than being invariant under all diffeomorphisms (Diff), its symmetry group consists of diffeomorphisms  that preserve the spacetime volume (transverse) and Weyl transformations (WTDiff). Weyl transverse gravity\footnote{The names Weyl transverse gravity and unimodular gravity are often used interchangeably. Many recent works prefer the term unimodular gravity~\cite{Carballo:2022,Alvarez:2023}. However, it also commonly refers to theories distinct from Weyl transverse gravity~\cite{Henneaux:1989,Padilla:2014,Bufalo:2015}. Therefore, we stick to the name Weyl transverse gravity for the purposes of the present paper.} originally emerged from the construction of a consistent theory for self-interacting gravitons~\cite{Alvarez:2006,Barcelo:2014,Alvarez:2016,Barcelo:2018,Carballo:2022}. It has been shown that two distinct theories can result from this construction, depending on the choice of the symmetry group. More precisely, these are the only two options with the maximum number of local symmetries, $D\left(D+1\right)/2$. Any other possibility involves gauge fixing. The standard Diff symmetry then leads to general relativity, whereas choosing the WTDiff symmetry yields Weyl transverse gravity.

In previous works, the similarity between gravitational dynamics implied by thermodynamics and Weyl transverse (or unimodular) gravity has been remarked~\cite{Tiwari:2006,Padmanabhan:2008,Alonso:2020,Alonso:2021}. However, these papers have not considered the self-consistency of the approach. Instead, they worked with a setup tailored for Diff-invariant gravitational dynamics and then pointed out the inconsistency of the result with general relativity.

Herein, we aim to provide a fully self-consistent analysis. First, as explained above, we are clear about the requirements we impose. We also check that these requirements are consistent with Weyl transverse gravity which we derive from them. In this regard, we verified the proportionality between entropy and area for Weyl transverse gravity in a previous work~\cite{Alonso:2022}. We also arguein another work~\cite{equivalence} that Weyl transverse gravity obeys the equivalence principle for self-gravitating bodies, being the only metric theory in four spacetime dimensions besides general relativity that does so. Moreover, we explicitly construct the local causal horizons in a way compatible with both Diff- and WTDiff-invariant spacetime geometry. In other words, our derivation remains agnostic about the symmetry group of gravitational dynamics and we only argue for WTDiff invariance based on the result we obtain.

In summary, we present a complete and self-contained argument for the recovery of Weyl transverse gravity from thermodynamics of local causal horizons. We do so without assuming in any way that gravitational dynamics \textit{emerges} as a thermodynamic limit of the behavior of some quantum degrees of freedom of the spacetime unrelated to the metric~\cite{Jacobson:1995ab,Padmanabhan:2010,Verlinde:2011}. We instead take a more modest position that thermodynamics \textit{encodes} all the relevant features of the gravitational dynamics, regardless of whether it is ultimately emergent or fundamental.

To complement our main result, we look at thermodynamics of local causal horizons from a different perspective. Here, we assume the WTDiff invariance from the beginning. We study a class of local, WTDiff-invariant purely metric theories, whose Lagrangians do not contain derivatives of the Riemann tensor. For these theories we show that their Wald entropy (derived in our previous works~\cite{Alonso:2022,Alonso:2022b}) encodes the gravitational equations of motion. This approach has been previously developed for the Diff-invariant case~\cite{Padmanabhan:2010,Jacobson:2012,Bueno:2017,Svesko:2018,Svesko:2019}. Showing that it also works for WTDiff-invariant theories primarily serves as a consistency check, although we also comment on some improvements over the Diff-invariant setup.
 
The paper is organized as follows. In section~\ref{WTG}, we recall the basics of Weyl transverse gravity and of more general WTDiff-invariant theories of gravity. Section~\ref{WTG from TD} contains the main part of the paper, i.e., the arguments for consistency of Weyl transverse gravity with thermodynamics of local causal horizons. To make our conclusions more robust, we discuss two different derivations of the equations governing gravitational dynamics, one based on tracking entropy flux across the local causal horizon, the other one on considering a small perturbation of the horizon away from the equilibrium state. In section~\ref{WTDiff from TD}, we derive the equations of motion for a class of WTDiff-invariant modified theories of gravity from their Wald entropy. Lastly, section~\ref{discussion} sums up our results.
	
Throughout this paper, we consider an arbitrary spacetime dimension $D$ (unless specified otherwise) and a metric signature $(-,+,...,+)$. We set $c=k_{\text{B}}=1$, but, to keep track of quantum and gravitational effects, we maintain $\hbar$ and $G$ explicit. We use lowercase Greek letters for spacetime indices and lowercase Latin letters for spatial indices. Other conventions follow~\cite{MTW}.

\section{Overview of Weyl transverse gravity and its generalizations}
\label{WTG}

The main aim of this paper is to discuss how thermodynamic arguments naturally lead to Weyl transverse gravity. To provide the necessary context for this discussion, we now briefly review the theory. A detailed exposition of Weyl transverse gravity and its generalizations can be found in~\cite{Carballo:2022}.

A key feature of Weyl transverse gravity is a non-dynamical  spacetime volume form determined by a strictly positive density $\omega$. Given a a metric $g_{\mu\nu}$ his volume measure allows us to define an auxiliary metric $\tilde g_{\mu\nu}$ whose determinant is equal to $-\omega^2$, i.e.,
\begin{equation}
\label{aux metric}
\tilde{g}_{\mu\nu}=\left(\sqrt{-\mathfrak{g}}/\omega\right)^{-2/D}g_{\mu\nu},
\end{equation}
where $\mathfrak{g}$ denotes the determinant of the dynamical metric $g_{\mu\nu}$. We also introduce a Levi-Civita connection $\tilde{\nabla}_{\mu}$ associated with $\tilde{g}_{\mu\nu}$ and the corresponding Riemann tensor $\tilde{R}^{\mu}_{\;\:\nu\rho\sigma}$. These quantities transform tensorially only under the subgroup of transverse diffeomorphisms which preserve the volume measure $\omega$, i.e.,
\begin{equation}
\delta_{\xi}\tilde{g}_{\mu\nu}=\tilde{\nabla}_{(\mu}\xi_{\nu)},\quad\tilde{\nabla}_{\mu}\xi^{\mu}=0,
\end{equation}
with $\xi^{\mu}$ being the infinitesimal diffeomorphism generator. They are unaffected by Weyl rescalings of the dynamical metric
\begin{equation}
\delta g_{\mu\nu}=e^{2\sigma}g_{\mu\nu},
\end{equation}
where $\sigma$ is an arbitrary function.

The simplest WTDiff-invariant gravitational action one can construct reads
\begin{equation}
\label{I WTG}
I_{\text{WTG}}=\frac{1}{16\pi G}\int_{\text{V}}\left(\tilde{R}+L_{\psi}\right)\omega\text{d}^{D}x,
\end{equation}
where $L_{\psi}$ denotes the matter action and $\tilde{R}=\tilde{g}^{\mu\nu}\tilde{R}_{\mu\nu}$ is the auxiliary scalar curvature (unless specified otherwise, we raise a lower the indices using the auxiliary metric $\tilde g_{\mu\nu}$). Varying action~\eqref{I WTG} with respect to the dynamical metric $g_{\mu\nu}$ yields the traceless equations of motion
\begin{equation}
\label{EoMs}
\tilde{R}_{\mu\nu}-\frac{1}{D}\tilde{R}\tilde{g}_{\mu\nu}=8\pi G\left(\tilde{T}_{\mu\nu}-\frac{1}{D}\tilde{T}\tilde{g}_{\mu\nu}\right),
\end{equation}
where we define the WTDiff-invariant energy-momentum tensor
\begin{equation}
\tilde{T}_{\mu\nu}=-2\frac{\partial L_{\psi}}{\partial\tilde{g}^{\mu\nu}}+L_{\psi}\tilde{g}_{\mu\nu}.
\end{equation}

While Diff invariance of gravitational dynamics directly implies the local energy-momentum conservation condition, $\nabla_{\nu}T_{\mu}^{\;\:\nu}=0$, this is not in general true for WTDiff-invariant theories. Nevertheless, WTDiff invariance of the matter action leads to a weaker condition~\cite{Alvarez:2013}
\begin{equation}
\label{div T}
\tilde{\nabla}_{\nu}\tilde{T}_{\mu}^{\;\:\nu}=\tilde{\nabla}_{\mu}\mathcal{J},
\end{equation}
where $\mathcal{J}$ is a scalar function. It is easy to see that if $\mathcal{J}\ne0$, then the energy-momentum tensor is not locally conserved (for a more detailed discussion of local energy-momentum non-conservation see, e.g.~\cite{Alonso:2021,Carballo:2022}). Nonetheless, the tensor
\begin{equation}
\label{T divergenceless}
\tilde{T}'_{\mu\nu}=\tilde{T}_{\mu\nu}-\mathcal{J}\tilde{g}_{\mu\nu},
\end{equation}
which will be relevant throughout, is indeed divergenceless.

Bianchi identities then allow us to rewrite the traceless equations of motion~\eqref{EoMs} in a divergenceless form
\begin{equation}
\label{Einstein-style}
\tilde{R}_{\mu\nu}-\frac{1}{2}\tilde{R}\tilde{g}_{\mu\nu}+\Lambda\tilde{g}_{\mu\nu}=8\pi G \tilde{T}'_{\mu\nu}.
\end{equation}
These equations have the same form as the Einstein equations with the divergenceless energy-momentum tensor $\tilde{T}'_{\mu\nu}$. We can see that the integration constant $\Lambda$ plays the role of the cosmological constant. In contrast with general relativity, $\Lambda$ has no connection with any fixed parameter present in the Lagrangian and is only defined on shell, having in principle different values for the various solutions of the theory. It has been shown that this behavior of the cosmological constant leads to its radiative stability in the effective field theory treatment of Weyl transverse gravity~\cite{Carballo:2015,Barcelo:2018,Carballo:2022}.

Weyl transverse gravity is also notable for incorporating the strong equivalence principle~\cite{equivalence}, which asserts that ``All test fundamental physics (including gravitational physics) is not affected locally by the presence of a gravitational field''~\cite{Casola:2015}. The only other metric theory of gravity in four spacetime dimensions compatible with this principle is general relativity (in higher dimensions, both Diff- and WTDiff-invariant Lovelock theories also incorporate it). This observation can be seen to follow from general relativity and Weyl transverse gravity being the only metric theories whose only degrees of freedom are those of a massless graviton~\cite{Barcelo:2014,Barcelo:2018,Carballo:2022}. It has been noted that the presence of any other propagating gravitational degrees of freedom tends to break the strong equivalence principle~\cite{Casola:2014}, although the Einstein equivalence principle (which does not concern gravitational test physics) remains valid even in these cases~\cite{Casola:2015}.

While we primarily focus on Weyl transverse gravity in the present work, we also show a thermodynamic derivation of equations of motion for a class of more general WTDiff-invariant theories of gravity in section~\ref{WTDiff from TD}. Specifically, we consider arbitrary gravitational Lagrangians constructed from the auxiliary metric, the auxiliary Riemann tensor (but not its derivatives) and minimally coupled matter fields. The most general such action reads
\begin{equation}
\label{I WTDiff}
I_{\text{WTDiff}}=\int_{\text{V}}L\left(\tilde{g}_{\mu\nu},\tilde{R}^{\mu}_{\;\:\nu\rho\sigma}\right)\omega\text{d}^{D}x+I_{\psi}.
\end{equation}
The corresponding traceless equations of motion are~\cite{Carballo:2022,Alonso:2022b}
\begin{equation}
\tilde{H}_{\mu\nu}-\frac{1}{D}\tilde{H}\tilde{g}_{\mu\nu}=8\pi G\left(\tilde{T}_{\mu\nu}-\frac{1}{D}\tilde{T}\tilde{g}_{\mu\nu}\right), \label{traceless WTDiff}
\end{equation}
where we defined the symmetric tensor
\begin{equation}
\tilde{H}_{\mu\nu}=16\pi G\left[\tilde{E}_{(\nu}^{\;\;\:\lambda\rho\sigma}\tilde{R}^{}_{\mu)\lambda\rho\sigma}-2\tilde{\nabla}_{\rho}\tilde{\nabla}_{\sigma}\tilde{E}_{(\mu\;\;\;\nu)}^{\;\;\:\rho\sigma}\right], \label{H}
\end{equation}
with the tensor $\tilde{E}_{\mu}^{\;\:\nu\rho\sigma}$ being the derivative of the Lagrangian with respect to the auxiliary Riemann tensor
\begin{equation}
\label{dL/dR}
\tilde{E}_{\mu}^{\;\:\nu\rho\sigma}=\frac{\partial L}{\partial\tilde{R}^{\mu}_{\;\:\nu\rho\sigma}}.
\end{equation}
For the special case of Weyl transverse gravity, $\tilde{H}_{\mu\nu}$ is simply the auxiliary Ricci tensor $\tilde{R}_{\mu\nu}$. In general, $\tilde{H}_{\mu\nu}$ depends on second derivatives of the auxiliary Riemann tensor, coming from the term $-2\tilde{\nabla}_{\rho}\tilde{\nabla}_{\sigma}\tilde{E}_{\mu\;\:\;\:\nu}^{\;\:\rho\sigma}$.

In the same way that Weyl transverse gravity represents a WTDiff-invariant alternative to general relativity, it has been shown that there exists a WTDiff-invariant theory corresponding to any Diff-invariant one~\cite{Carballo:2022}. Such pairs of corresponding theories have the same classical dynamics, except for the different behavior of $\Lambda$. Likewise, for every WTDiff-invariant theory that incorporates local energy-momentum conservation, there exists a Diff-invariant theory equivalent to it in this sense.

\section{Weyl transverse gravity from thermodynamics}
\label{WTG from TD}

We now discuss how the gravitational dynamics follows from thermodynamics of local causal horizons. Our aim is to show that the resulting dynamics is distinct from general relativity and instead equivalent to Weyl transverse gravity. To that end, we derive the equations governing gravitational dynamics from the minimal thermodynamic setup, involving as few assumptions as possible. In fact, as we foreshadowed in the introduction, the only nontrivial requirements we impose are that a local causal horizon possesses entropy proportional to its area (regardless of its microscopic origin) and that the strong equivalence principle holds. The equations for gravitational dynamics are then encoded in an equilibrium relation applied to the Clausius entropy flux across the horizon and the corresponding changes in the horizon area. We carry out this derivation in subsection~\ref{physical process Weyl transverse gravity}.

In subsection~\ref{entanglement equilibrium}, we discuss an independent derivation which considers a small perturbation away from the equilibrium state and the corresponding changes in entropy. This approach involves an extra assumption that both the horizon and matter entropy can be interpreted in terms of quantum von Neumann entropy~\cite{Jacobson:2015}. While this assumption somewhat lessens the generality of the derivation, it allows us to obtain the semiclassical equations governing the gravitational dynamics, which couple the classical spacetime curvature to the quantum expectation value of the energy-momentum tensor.

For both derivations, we want to decide whether the resulting gravitational dynamics corresponds to general relativity or Weyl transverse gravity. Therefore, we remain agnostic as to whether the local causal horizon is defined with respect to the dynamical metric $g_{\mu\nu}$ (as it would be for general relativity), or the auxiliary metric $\tilde{g}_{\mu\nu}$ (for Weyl transverse gravity). To take into account both possibilities in our notation, we use hatted quantities such as $\hat{g}_{\mu\nu}$ (which will be used to raise and lower the indices), $\hat{\mathcal{A}}$, $\hat{T}_{\mu\nu}$ and so on throughout this section. These can either mean the Diff-invariant expressions, or the corresponding WTDiff-invariant ones. In this way, we avoid repeating the analysis twice.

\begin{figure}[tbp]
    \centering
    \includegraphics[width=.45\textwidth,origin=c,trim={4.1cm 1.3cm 20.6cm 1.8cm},clip]{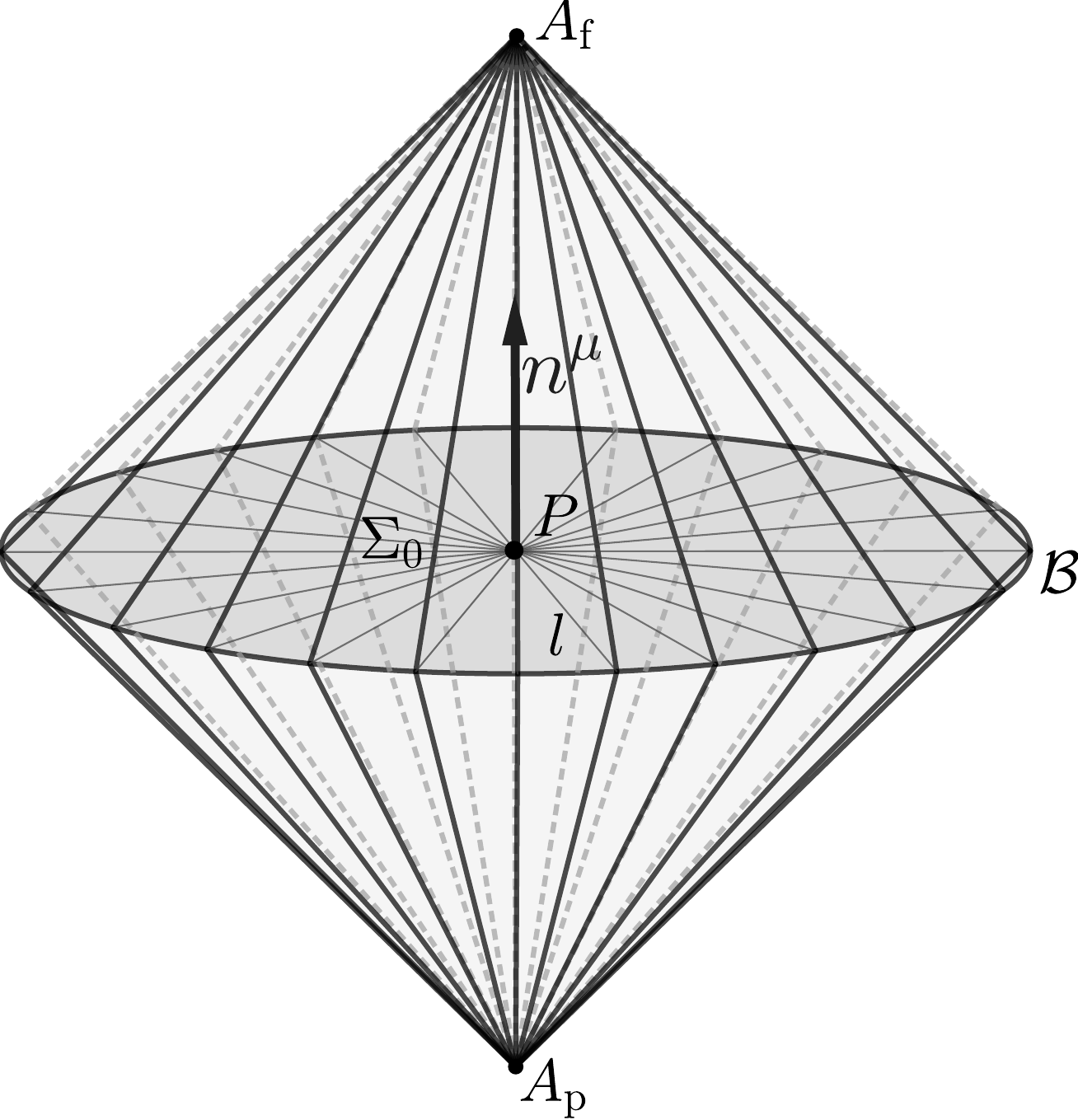}
    \caption{\label{diamond} A causal diamond centered in a flat spacetime point $P$. We suppress $D-3$ angular coordinates. The unit, future-directed timelike vector $n^{\mu}$ defines the local direction of time. The diamond's base is a $\left(D-1\right)$-dimensional spacelike ball $\Sigma_0$ centered in $P$ and of radius $l$. Its boundary $\mathcal{B}$ is an approximate $\left(D-2\right)$-sphere. The null generators of the diamond's boundary are depicted by lines starting in the diamond's past apex $A_{\text{p}}$ ($t=-l$) and ending in the future apex $A_{\text{f}}$ ($t=l$). The ball $\Sigma_0$ lies at the intersection of a future light cone starting in $A_{\text{p}}$ and a past light cone ending in $A_{\text{f}}$.}
\end{figure}

While the seminal thermodynamic derivations worked with local Rindler horizons~\cite{Jacobson:1995ab,Eling:2006,Chirco:2010,Padmanabhan:2010}, we instead choose causal diamonds (see, e.g.~\cite{Jacobson:2015,Carroll:2016,Svesko:2018,Svesko:2019,Jacobson:2019} for the advantages of this choice). In flat spacetime, a causal diamond is unambiguously defined as the domain of dependence of a spacelike $\left(D-1\right)$-dimensional ball. Then, it is fully specified by the center of the ball $P$, the ball's geodesic radius $l$ and the local choice of the direction of time, given by a unit timelike vector $\hat{n}^{\mu}$. We display the construction in figure~\ref{diamond}.

Causal diamonds are invariant under transformations generated by the following conformal Killing vector
\begin{equation}
\label{conformal Killing}
\zeta^{\mu}=C\left[\left(l^2-t^2-r^2\right)\left(\partial_t\right)^{\mu}-2rt\left(\partial_r\right)^{\mu}\right],
\end{equation}
where $r$ stands for the radial geodesic distance from point $P$, $t$ is the time coordinate measured along vector $\hat{n}^{\mu}$, and $C$ denotes an arbitrary constant determining the normalization of $\zeta^{\mu}$. The conformal Killing vector $\zeta^{\mu}$ is null on the diamond's boundary and vanishes at $\mathcal{B}$. Thence, the boundary represents a bifurcate conformal Killing horizon and the $\left(D-2\right)$-sphere $\mathcal{B}$ its bifurcation surface.

In a generic curved spacetime, causal diamonds can only be constructed locally, with their size parameter $l$ being much smaller than the local curvature length scale (inverse of the square root of the largest eigenvalue of the Riemann tensor). We also require $l$ to be much larger than the Planck length $l_{\text{P}}=\sqrt{G\hbar}$, as there exist strong indications that the standard description of the spacetime breaks down at this length scale~\cite{Mead:1964,Garay:1994en,Hossenfelder:2013}. Even if $l$ obeys both conditions, we have several non-equivalent ways to extend the definition of a local causal diamond (LCD) to a curved spacetime~\cite{Wang:2019}. The particulars of the construction of an LCD are not relevant for our conclusions. We simply need a locally constructed causal horizon whose spatial cross-section is an approximate sphere. That being said, some definitions of LCDs are especially well suited for the deriving the equations governing gravitational dynamics as we discuss in the following.

\subsection{Minimal thermodynamic setup: physical process approach}
\label{physical process Weyl transverse gravity}

We start by discussing the physical process derivation of the equations governing the gravitational dynamics from thermodynamics. The method we use further builds upon the framework previously explored in the literature~\cite{Svesko:2018,Svesko:2019,Alonso:2021}.

The basic idea of the derivation is simple. It has been shown by a number of methods that any causal horizon possesses entropy proportional to the area of its spatial cross-section~\cite{Bekenstein:1973,Sorkin:1986,Srednicki:1993,Jacobson:2003,Solodukhin:2011,Chakraborty:2016,Jacobson:2019,Jacobson:2023,Jacobson:2023b,Speranza:2023}, i.e.,
\begin{equation}
\label{area entropy}
S=\eta\hat{\mathcal{A}}
\end{equation}
where $\eta$ is a universal constant, provided that the strong equivalence principle holds~\cite{Chirco:2010,Jacobson:2015}. Notably, this result for entropy is independent of gravitational dynamics. Various microscopic interpretations of this entropy have been put forward (essentially mirroring those proposed for black hole entropy). For the purposes of this subsection, we remain agnostic as to which of these interpretations is correct. We only require that equation~\eqref{area entropy} holds. In particular, our conclusions do not in any way rely on the entanglement interpretation of the entropy~\cite{Sorkin:1986,Srednicki:1993,Belgiorno:1996,Liberati:1997,Solodukhin:2010,Solodukhin:2011}. Interested readers can find a short review of arguments for assigning entropy to local causal horizons in appendix~\ref{lcd entropy}.

As the LCD evolves in time, its horizon expands, leading to a change of its entropy, $\Delta S=\eta\Delta\hat{\mathcal{A}}$ (recall that $\eta$ is a universal constant). At the same time, matter crosses the LCD's horizon can be interpreted as a heat flux $\Delta Q$~\cite{Jacobson:1995ab,Baccetti:2013ica}. Now consider a uniformly accelerating observer moving inside the LCD. This observer perceives both the changes in the horizon area and the matter flux. Furthermore, due to the Unruh effect, they see the local Minkowski vacuum inside the LCD as a thermal state at the Unruh temperature $T_{\text{U}}$ proportional to their acceleration~\cite{Fulling:1973,Bisognano:1976,Unruh:1976,Barbado:2012,Baccetti:2013ica,Perche:2021,Perche:2022b} (provided that the acceleration is much larger than $1/l$, so that the observer's detector has enough time to thermalize). A reversible thermodynamic process perceived by such an observer in the limit of infinite acceleration then obeys the Clausius relation
\begin{equation}
\label{Clausius}
\eta\Delta\hat{\mathcal{A}}=\frac{\Delta Q}{T_{\text{U}}},
\end{equation}
where the the right hand side corresponds to the matter Clausius entropy flux $\Delta S_{\text{C}}=\Delta Q/T_{\text{U}}$ across the given slice of the LCD's horizon~\cite{Jacobson:1995ab,Baccetti:2013ica,Alonso:2021}. Naturally, in the limit of infinite acceleration, $T_{\text{U}}$ diverges. However, $\Delta Q$ diverges in this limit at the same rate as $T_{\text{U}}$, so that the ratio $\Delta Q/T_{\text{U}}$ remains finite and well defined~\cite{Baccetti:2013ica}. In the following, we show that relation~\eqref{Clausius} is equivalent to the traceless Einstein equations.

\begin{figure}[tbp]
\centering
\includegraphics[width=8.5cm,origin=c,trim={0.85cm 0.75cm 0.85cm 0.6cm},clip]{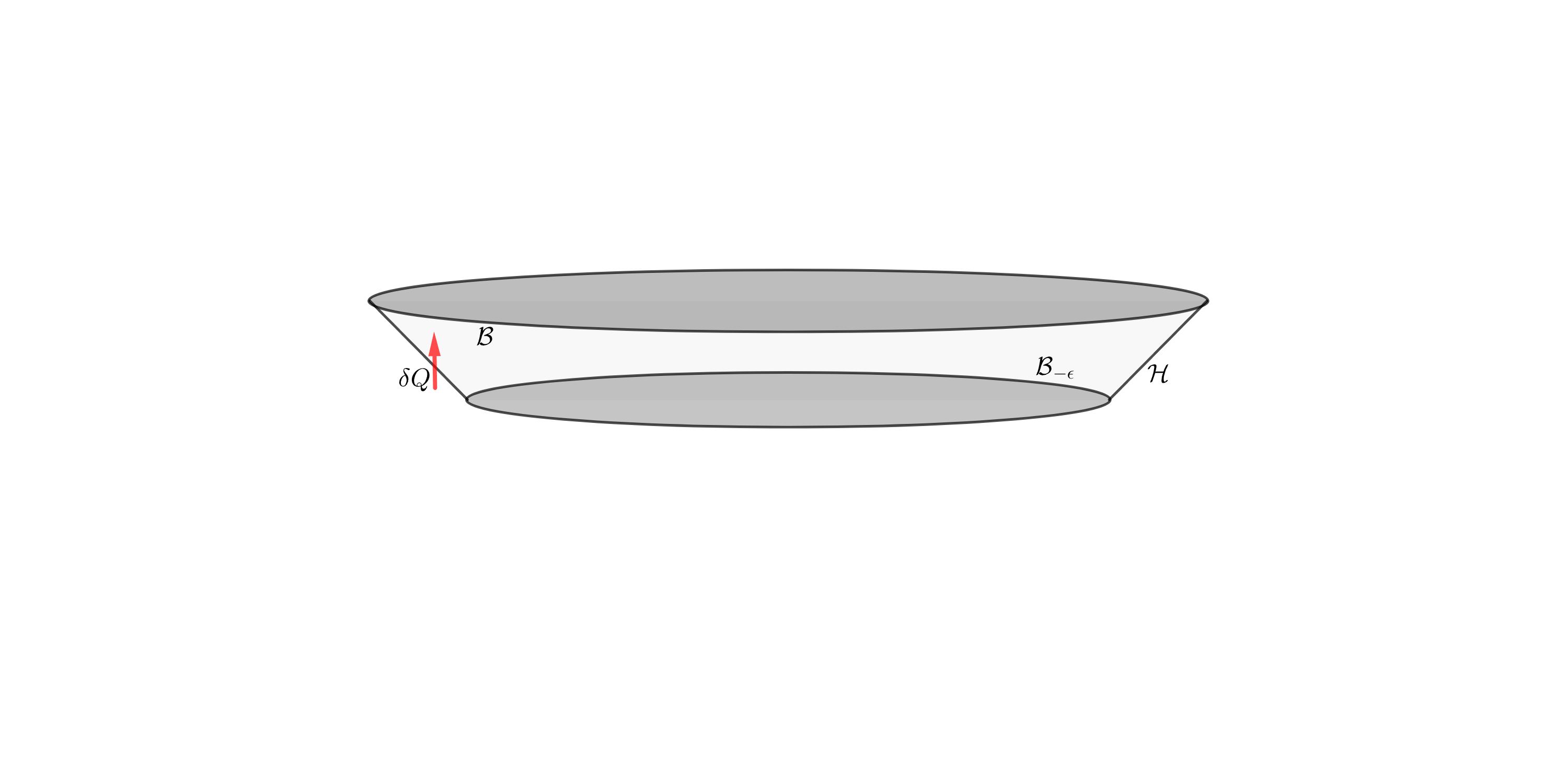}
\caption{\label{slice}{A sketch of a slice of a light-cone cut LCD. We denote the spatial cross-sections at times $t=-\epsilon$ and $t=0$ by $\mathcal{B}_{-\epsilon}$ and $\mathcal{B}$, respectively. The red arrow depicts the physical heat flux $\delta Q$ across the slice's null boundary $\mathcal{H}$.}}
\end{figure}

The idea is to study the change in the entropy of an LCD between two instances of time $t=-\epsilon$ and \mbox{$t=0$}, where $\epsilon$ is taken to be much smaller than the size parameter $l$, $\epsilon\ll l$. While this requirement is not strictly necessary~\cite{Alonso:2020}, it simplifies the calculations by allowing us to drop the subleading terms in~$\epsilon$~\cite{Svesko:2019}. Hence, we work with a slice of the LCD's past horizon bounded by the approximate $\left(D-2\right)$-sphere $\mathcal{B}_{-\epsilon}$ at $t=-\epsilon$ and by the approximate $\left(D-2\right)$-sphere $\mathcal{B}$ at $t=0$ (see figure~\ref{slice}).
	
In the limit of infinite acceleration (i.e., when the surface swept out by the accelerating observer approaches the horizon), the right hand side of equation~\eqref{Clausius} evaluated between times $t_1$ and $t_2$ equals~\cite{Baccetti:2013ica,Alonso:2020}
\begin{equation}
\Delta S_{\text{C}}=-\epsilon^2\frac{\pi\Omega_{D-2}l^{D-2}}{\hbar\left(D-1\right)}\hat{T}_{\mu\nu}\left(D\hat{n}^{\mu}\hat{n}^{\nu}+\hat{g}^{\mu\nu}\right)+O\left(\epsilon^2l^D\right), \label{dSC}
\end{equation}
where $\Omega_{D-2}$ denotes the area of a unit $\left(D-2\right)$-sphere. We expanded the result in the small time interval $\epsilon$, discarding all the subleading $O\left(\epsilon^3\right)$ terms. The negligible correction term $O\left(\epsilon^2l^D\right)$ appears due to approximating the energy-momentum tensor by its value in $A_{\text{p}}$ and due to disregarding the curvature-dependent effects.

Next, we need to evaluate the change in the horizon area $\Delta\hat{\mathcal{A}}$. In this case, the curvature effects contribute already at the order $O\left(\epsilon^2l^{D-2}\right)$ and cannot be neglected. Since the geometry of the LCD is not unique in a generic curved spacetime, we require a way to fix it. To compute $\Delta\hat{\mathcal{A}}$, we must fully specify the geometry of a slice of the LCD's past null boundary. A definition of an LCD satisfying this requirement is known as the light-cone cut LCD~\cite{Wang:2019}. To construct it we begin at a point $A_{\text{p}}$, the past apex of the eventual diamond (see figure~\ref{diamond}). We fix the unit timelike vector $\hat{n}^{\mu}$ as the local direction of time and take the future directed null vector fields $\hat{k}_{-}^{\mu}$ at $A_{\text{p}}$ normalized so that $\hat{n}_{\mu}\hat{k}_{-}^{\mu}=-1$. We then construct the past boundary of the LCD as a congruence of null wordlines tangent to $\hat{k}_{-}^{\mu}$. The spacelike cross-section of this congruence at the affine parameter length $l$ measured along $\hat{k}_{-}^{\mu}$ corresponds to an approximate $\left(D-2\right)$-sphere $\mathcal{B}$, whose interior, an approximate $\left(D-1\right)$-dimensional spacelike ball $\Sigma_0$, is the base of the light-cone cut LCD. We call the center of the ball $P$.

The construction of the future (contracting) part of the light-cone cut LCD is, as mentioned above, not needed to be fixed for our purposes. The most straightforward option would be to specify the future-directed null vector fields $\hat{k}_{+}^{\mu}$ on $\mathcal{B}$. We can define them so that \mbox{$\hat{n}_{\mu}\hat{k}_{+}^{\mu}=-1$} and, denoting the projection of $k_{-}^{\mu}$ on the surface orthogonal to $\hat{n}^{\mu}$ by $\hat{m}^{\mu}$, the same projection of $\hat{k}_{+}^{\mu}$ is $-\hat{m}^{\mu}$. In other words, we choose the congruence with a negative expansion. Then, the congruence of the null wordlines tangent to $k_{+}^{\mu}$ forms the future boundary of the causal diamond.

For a light-cone cut LCD we can easily compute the difference in the area of the boundary's spatial (i.e., orthogonal to the vector field $\hat{n}^{\mu}$) cross-sections at different times, in our case at $t=-\epsilon$ and $t=0$. The simplest way to do it is by considering the expansion of the congruence of the null boundary generators $\hat{k}^{\mu}_{-}$, i.e., $\theta=\hat{\nabla}_{\mu}\hat{k}^{\mu}_{-}$. By definition of the expansion, it then holds for the change of area between times $t=-\epsilon$ and $t=0$~\cite{Raychaudhuri:1955,Jacobson:1995ab,Wang:2019}
\begin{equation}
\label{area change}
\Delta\hat{\mathcal{A}}=\int_{-\epsilon}^{0}\text{d}s\int\text{d}^{D-2}\hat{\mathcal{A}}\,\theta,
\end{equation}
with $s$ being the null parameter along the horizon generators. The evolution of $\theta$ obeys the Raychaudhuri equation~\cite{Raychaudhuri:1955}
\begin{equation}
\label{Raychaudhuri}
\dot{\theta}=-\frac{1}{D-2}\theta^2-\sigma^2-\hat{R}_{\mu\nu}\hat{k}_{-}^{\mu}\hat{k}_{-}^{\nu},
\end{equation}
where we introduced the shorthand $\dot{\theta}=\text{d}\theta/\text{d}s$. The second term on the right hand side \mbox{$\sigma^2=\sigma_{\mu\nu}\sigma^{\mu\nu}$} corresponds to the shear of the congruence
\begin{equation}
\sigma_{\mu\nu}=\hat{h}_{\mu}^{\;\:\lambda}\hat{h}_{\nu}^{\;\:\rho}\hat{\nabla}_{(\lambda\vert}\hat{k}_{-\vert\rho)}-\frac{1}{D-2}\hat{\nabla}_{\rho}\hat{k}_{-}^{\rho}\hat{h}_{\mu\nu},
\end{equation}
with $\hat{h}_{\mu\nu}$ being the induced metric on the null boundary (the shear does not depend on its precise choice). The twist of the congruence vanishes because it generates a surface. The shear tensor evolves according to the following equation
\begin{align}
\label{Sigma}
&\nonumber \dot{\sigma}_{\mu\nu}=-\frac{2}{D-2}\theta\sigma_{\mu\nu}-\sigma_{\mu\lambda}\sigma_{\nu}^{\;\:\lambda}+\frac{1}{D-2}\sigma^2\hat{h}_{\mu\nu} \\
&-\hat{C}_{\lambda\rho\sigma\tau}\hat{k}_{-}^{\lambda}\hat{k}_{-}^{\sigma}\hat{h}_{\mu}^{\rho}\hat{h}_{\nu}^{\tau}+\frac{1}{2}\left(\hat{h}_{\mu\lambda}\hat{h}_{\nu\rho}-\frac{1}{D-2}\hat{h}_{\mu\nu}\hat{h}_{\lambda\rho}\right)\hat{R}^{\lambda\rho},
\end{align}
where $\hat{C}_{\lambda\rho\sigma\tau}$ is Weyl curvature tensor. The horizon of an LCD in flat spacetime has an identically vanishing shear. However, its expansion equals \mbox{$\theta_{\text{flat}}\left(s\right)=\left(D-2\right)/\left(l+s\right)$}~\cite{Wang:2019}. The curvature-dependent terms in the evolution equations for $\theta$ and $\sigma_{\mu\nu}$ do not contain any further terms inversely proportional to $s$. Thence, we can expand $\theta$ and $\sigma_{\mu\nu}$ in powers of $s$ in the following way (note that $-\epsilon\le s\le0$)
\begin{align}
\theta&=\theta_{\text{flat}}+\theta_{(0)}+s\theta_{(1)}+O\left(s^2\right), \label{Rauch} \\
\sigma_{\mu\nu}&=\sigma_{(0)\mu\nu}+s\sigma_{(1)\mu\nu}+O\left(s^2\right). \label{shear}
\end{align}

In general, $\theta_{(0)}$ and $\sigma_{(0)\mu\nu}$ represent $D\left(D+1\right)/2$ arbitrary functions. However, it has been shown that one can refine the construction of a light-cone cut LCD~\cite{Svesko:2018,Svesko:2019}. While the motivation in that case has been the use of the conformal Killing identity rather than the Raychaudhuri equation, it also involves fixing $D\left(D+1\right)/2$ arbitrary functions. Therefore, translating the results of that analysis to the language of the Rauchaudhuri equation implies that we are free to set up our LCD so that $\theta_{(0)}=\sigma_{(0)\mu\nu}=0$. Our choice here differs from the one made in reference~\cite{Wang:2019}, which sets $\theta=\sigma_{\mu\nu}=0$ at the past apex $A_{\text{p}}$. In principle, we might also keep $\theta_{(0)}$ and $\sigma_{(0)\mu\nu}$ arbitrary. In that case, a previous analysis suggest that they would correspond to non-equilibrium entropy production~\cite{Chirco:2010}. The outcome of the derivation remains the same regardless of whether we keep $\theta_{(0)}$ and $\sigma_{(0)\mu\nu}$ arbitrary or not. Nevertheless, for the sake of clarity, we proceed assuming that we define our light-cone cut LCD so that $\theta_{(0)}=\sigma_{(0)\mu\nu}=0$.

Plugging the ansatze~\eqref{Rauch} and~\eqref{shear} for the expansion and the shear into the evolution equations~\eqref{Raychaudhuri} and~\eqref{Sigma} yields the following solution for the expansion
\begin{equation}
\theta=\theta_{\text{flat}}-s\hat{R}_{\mu\nu}\hat{k}_{-}^{\mu}\hat{k}_{-}^{\nu}+O\left(s^2\right).
\end{equation}
The flat spacetime expansion $\theta_{\text{flat}}$ is clearly not related to the spacetime curvature. Moreover, the area change proportional to $\theta_{\text{flat}}$  occurs even in vacuum, with no Clausius entropy flux across the horizon $\mathcal{H}$. Thence, we split the area change in two parts, one proportional to $\theta_{\text{flat}}$, the other to $\theta_{\text{curv}}=-s\hat{R}_{\mu\nu}\hat{k}_{-}^{\mu}\hat{k}_{-}^{\nu}$. Only the latter part can correspond to equilibrium change in the entropy of the horizon that is balanced by a matter entropy flux (see also~\cite{Svesko:2018,Svesko:2019} for an alternative interpretation of $\theta_{\text{flat}}$ in terms of an irreversible thermodynamic process).

To compute the equilibrium change in area corresponding to $\theta_{\text{curv}}$ between times $t=-\epsilon$ and $t=0$ we now simply need to substitute $\theta_{\text{curv}}$ into equation~\eqref{area change}, obtaining
\begin{align}
\nonumber \Delta\hat{\mathcal{A}}_{\text{curv}}=&-\frac{\epsilon^2}{2}\int\hat{R}_{\mu\nu}\hat{k}_{-}^{\mu}\hat{k}_{-}^{\nu}\text{d}^{D-2}\hat{\mathcal{A}}+O\left(\epsilon^2l^D\right), \\
=&-\epsilon^2\frac{\Omega_{D-2}l^{D-2}}{2\left(D-1\right)}\hat{R}_{\mu\nu}\left(D\hat{n}^{\mu}\hat{n}^{\nu}+\hat{g}^{\mu\nu}\right)+O\left(\epsilon^2l^D\right), \label{area expansion}
\end{align}
where we approximate the Ricci tensor by its value in $A_{\text{p}}$, leading to an $O\left(\epsilon^2l^D\right)$ error. Further errors of the same order appear due to neglecting the subleading curvature-dependent terms. The equilibrium change of the horizon entropy then equals $\Delta S_{\text{curv}}=\eta\Delta\hat{\mathcal{A}}_{\text{curv}}$.
	
The total equilibrium change in entropy vanishes, which implies $\Delta S_{\text{curv}}+\Delta S_{\text{C}}=0$. After some straightforward simplifications, this condition becomes
\begin{equation}
\label{precursor}
\left(\hat{R}_{\mu\nu}-\frac{1}{D}\hat{R}\hat{g}_{\mu\nu}-\frac{2\pi}{\eta\hbar}\hat{T}_{\mu\nu}+\frac{2\pi}{\eta\hbar}\frac{1}{D}\hat{T}\hat{g}_{\mu\nu}\right)\hat{n}^{\mu}\hat{n}^{\nu}=0,
\end{equation}
valid at the given point point $A_{\text{p}}$. The construction of a light-cone cut causal LCD and derivation of equation~\eqref{precursor} can be performed for every unit, timelike vector field $\hat{n}^{\mu}$ defined at $A_{\text{p}}$. Since equation~\eqref{precursor} holds for an arbitrary unit, timelike vector, it implies (see the proof in appendix~\ref{n-dependence})
\begin{equation}
\label{precursor 2}
\hat{R}_{\mu\nu}-\frac{1}{D}\hat{R}\hat{g}_{\mu\nu}\bigg\vert_{A_{\text{p}}}=\frac{2\pi}{\eta\hbar}\left(\hat{T}_{\mu\nu}-\frac{1}{D}\hat{T}\hat{g}_{\mu\nu}\right)\bigg\vert_{A_{\text{p}}}.
\end{equation}
We assume that the strong equivalence principle holds. Consequently, $\eta$ is a universal constant. If that were not the case, measuring entropy of two identical test black holes at different spacetime points could distinguish them, breaking the equivalence principle for self-gravitating test particles. Furthermore, equations~\eqref{precursor 2} can be derived at any regular spacetime point and have the same form at every point $A_{\text{p}}$. Finally, by considering the Newtonian limit of equations~\eqref{precursor 2}, we may \textit{define} the Newton gravitational constant in terms of $\eta$, i.e., $G=1/\left(4\hbar\eta\right)$~\cite{Chirco:2010,Jacobson:2015}. The horizon entropy \mbox{$S=\eta\hat{\mathcal{A}}$} then agrees with the Bekenstein entropy prescription \mbox{$S_{\text{B}}=\hat{\mathcal{A}}/4\hbar G$}.
	
In total, we have derived the following traceless equations governing gravitational dynamics
\begin{equation}
\label{traceless phys}
\hat{R}_{\mu\nu}-\frac{1}{D}\hat{R}\hat{g}_{\mu\nu}=8\pi G\left(\hat{T}_{\mu\nu}-\frac{1}{D}\hat{T}\hat{g}_{\mu\nu}\right).
\end{equation}
Taking the divergence of equations~\eqref{traceless phys} and invoking Bianchi identities implies $\hat{\nabla}_{\nu}\hat{T}_{\mu}^{\;\:\nu}=\hat{\nabla}_{\mu}\mathcal{J}$, where $\mathcal{J}$ is an arbitrary function. Then, we obtain the following divergenceless equations
\begin{equation}
\label{divergenceless phys}
\hat{R}_{\mu\nu}-\frac{1}{2}\hat{R}\hat{g}_{\mu\nu}+\Lambda\hat{g}_{\mu\nu}=8\pi G\hat{T'}_{\mu\nu},
\end{equation}
where $\Lambda$ is an arbitrary integration constant and \mbox{$\hat{T'}_{\mu\nu}=\hat{T}_{\mu\nu}-\mathcal{J}\hat{g}_{\mu\nu}$}. We stress that the equations for gravitational dynamics we obtain from thermodynamics are the traceless ones~\eqref{traceless phys}, equations~\eqref{divergenceless phys} only arise by integrating them. Therefore, the cosmological constant $\Lambda$ is only meaningful on shell and in principle varies between the solutions.

Up to this point, we have been agnostic about the local symmetries of our setup. Now, upon deriving the equations for gravitational dynamics, we are in a position to discuss the possible symmetry groups. Equations~\eqref{divergenceless phys} contain the metric $\hat{g}_{\mu\nu}$ as the only gravitational degree of freedom. Since $\hat{g}_{\mu\nu}$ is a symmetric tensor, it can have at most  $D\left(D+1\right)/2$ local symmetries. Assuming that we do not introduce any gauge fixing, we have only two choices compatible with the strong equivalence principle; the Diff and the WTDiff groups. We can understand the privileged position of these two groups in the following way: the strong equivalence principle can only be incorporated in theories that have just the two propagating degrees of freedom, the ones associated with a massless graviton~\cite{Casola:2014,Casola:2015}. Let us suppose that we write down a representation for a massless graviton with two physical polarizations in flat spacetime. We want to describe such a graviton by a symmetric, rank $2$ tensor that carries the maximum amount of gauge symmetry, i.e., we do not wish to introduce any gauge fixing. Then, the gauge group can be either Diff or WTDiff and the corresponding linearized action corresponds either to general relativity or to Weyl transverse gravity, respectively~\cite{Alvarez:2006}. This perspective singles out general relativity and Weyl transverse gravity as the only gravitational theories with two propagating degrees of freedom (the massless graviton) and the maximum amount of gauge symmetry. It then follows that these two theories are the only ones compatible with the strong equivalence principle.

We cannot directly study the local symmetries using thermodynamic reasoning, but we are nevertheless able to argue for its consistency with the WTDiff group. The key point is that local causal horizons (regardless of their specific realization) are insensitive to the overall conformal factor of the metric, which does not change the causal structure. Consequently, only the traceless equations~\eqref{traceless phys} directly follow from the equilibrium condition~\eqref{precursor}. These indeed fix the dynamical metric $g_{\mu\nu}$ only up to the overall conformal factor~\cite{Hawking:1973}. However, they suffice to recover the $D\left(D-1\right)/2$ components of the WTDiff-invariant auxiliary metric $\tilde{g}_{\mu\nu}$. Equations~\eqref{traceless phys} are then fully consistent (together with the matter equations of motion, of course) without any further assumptions if and only if we write them in terms of the WTDiff-invariant auxiliary tensors. Therefore, they coincide with the equations of motion of Weyl transverse gravity~\eqref{EoMs}.

Moreover, the gravitational equations we derived are encoded in the change of the horizon entropy. Then, shifting entropy by a universal constant has no effect on the gravitational dynamics. In a previous work, we have shown on the example of a de Sitter horizon that its entropy is indeed only defined up to a universal constant in Weyl transverse gravity~\cite{Alonso:2022}. However, we apparently have no freedom to similarly shift the horizon entropy in general relativity.

Finally, suppose we want to write an effective Lagrangian that implies the traceless equations~\eqref{traceless phys} we derived from thermodynamics. If gravity is a fundamental interaction, such a Lagrangian should exist and play an important role in the quantum theory (as it does in loop quantum gravity or path integral quantum gravity). If gravity is emergent, we still find it reasonable to expect that we can write some effective classical Lagrangian for it. However, there exists no Diff-invariant metric (with or without extra fields) Lagrangian whose equations of motion are the traceless equations~\eqref{traceless phys}~\cite{Garcia:2023} (although non-metric proposals for the variational principle have been put forward~\cite{Padmanabhan:2010,Montesinos:2023}). At the same time, equations~\eqref{traceless phys} coincide with the equations of motion obtained from the Weyl transverse gravity Lagrangian~\eqref{I WTG}. Therefore, assuming that equations~\eqref{traceless phys} we derived from thermodynamics are the Euler-Lagrange equations of some metric action, we are uniquely led to Weyl transverse gravity.
	
In principle, we have  no \textit{a priori} reason to expect that thermodynamics (together with the equivalence principle) recovers all the information about gravitational dynamics. Perhaps some information, i.e., the local energy-momentum conservation \textit{and} a fixed off-shell value of the cosmological constant indeed needs to be added, allowing us to obtain a Diff-invariant description of gravity. Nevertheless, the apparent strong connection between thermodynamics and gravity makes it worthwhile to ask what happens if they are in fact fully equivalent. And pursuing this question leads us directly to Weyl transverse gravity. Remarkably, this theory is also supported by a number of arguments completely independent of thermodynamics. First, the field theoretic approach to gravity~\cite{Alvarez:2006,Barcelo:2014,Carballo:2022} singles out Weyl transverse gravity and general relativity, putting both theories on the same ground. Furthermore, we have shown that Weyl transverse gravity and general relativity are apparently the only two theories of gravity in four spacetime dimensions satisfying the strong equivalence principle~\cite{equivalence}. Moreover, in contrast to general relativity, Weyl transverse gravity also offers a robust solution to the problem of the vacuum energy contributions to the cosmological constant~\cite{Carballo:2015,Barcelo:2018}.
	
In conclusion, thermodynamics of spacetime encodes the traceless equations of motion of Weyl transverse gravity~\eqref{EoMs}. All the hatted quantities we used throughout the derivation ought to be understood as being WTDiff-invariant, i.e., defined with respect to the auxiliary metric tensor $\tilde{g}_{\mu\nu}=\hat{g}_{\mu\nu}$.

\subsection{Semiclassical dynamics: entanglement equilibrium approach}
\label{entanglement equilibrium}

To strengthen our case for Weyl transverse gravity, we also analyze its consistence with the entanglement equilibrium approach to derive the equations governing the gravitational dynamics. This derivation is independent of the physical process one that we studied in the previous subsection. The precise relation between both approaches is subtle and not yet completely understood. We briefly discuss it in subsection~\ref{comparison} after showing how both derivations work in the context of modified theories of gravity.

The entanglement equilibrium approach phrases the local equilibrium conditions fully in terms of perturbations of the total quantum von Neumann entropy~\cite{Jacobson:2015}. There are two contributions to the total entropy perturbation, one coming from the matter fields and the other one from the entanglement entropy of the horizon, proportional to its area $S=\eta\hat{\mathcal{A}}$.

On the one hand, this method has the drawback of relying on a specific interpretation of the horizon entropy.  On the other hand, it has the advantage of using the same (von Neumann) definition both for entropy of the horizon and of matter. The entanglement equilibrium approach also becomes particularly natural in the AdS/CFT paradigm~\cite{Lashkari:2014,Faulkner:2014,Faulkner:2017}, in which the entropy of a causal horizon can be manifestly accounted for in terms of von Neumann entropy (via the Ryu-Takayanagi formula~\cite{Ryu:2006}). However, we instead consider the completely general spacetime setting, following the seminal paper~\cite{Jacobson:2015}. 

Our starting point is an LCD's spatial cross-section $\Sigma_0$ at time $t=0$ (a $\left(D-1\right)$-dimensional ball) in its equilibrium state. It has been proposed that the appropriate local equilibrium state corresponds to a vacuum, maximally symmetric spacetime with curvature $\hat{G}^{\text{MSS}}_{\mu\nu}=-\lambda\hat{g}^{\text{MSS}}_{\mu\nu}$~\cite{Jacobson:2015} (so that, if this spacetime was a solution of the equations of motion of either general relativity or Weyl transverse gravity, $\lambda$ would correspond to the cosmological constant). In principle, the local value of $\lambda$ depends both on the position of the LCD and on its size parameter $l$. The equilibrium condition involves both the background and the value of the possible perturbations. This means that only a certain background corresponds to the equilibrium state. More explicitly, the equilibrium condition $\delta S=0$ ($S$ being the total entropy of the LCD) not only constrains the allowed perturbations, but also provides the value of $\lambda$ corresponding to the equilibrium state. In other words, $\lambda$ is determined by the perturbation of the corresponding state being isentropic, i.e., the net change of the LCD's entropy due to perturbation vanishing to the first order. We determine the equilibrium $\lambda$ only at the end of our derivation, keeping its value unspecified for the time being.

Taking the LCD in this equilibrium state, we introduce a small arbitrary perturbation of the metric $\delta\hat{g}_{\mu\nu}$ and of the matter fields, leading to a non-zero perturbation of the expectation value of the energy-momentum tensor, $\delta\langle\hat{T}^{\mu\nu}\rangle$. Once again, we proceed without presuming anything about the local symmetries of our setup. Since choosing the WTDiff symmetry group would also generically imply variations of the cosmological constant, we further allow the possibility of a small variation of the local curvature, $\delta\lambda$ (that would vanish in the Diff-invariant case).

For a perturbed equilibrium state, the total entropy perturbation by definition vanishes to  leading order~\cite{Jacobson:2015}, i.e.,
\begin{equation}
\label{equilibrium}
\eta\delta\hat{\mathcal{A}}+\delta S_{\text{vN}}=0,
\end{equation}
where the first term corresponds to the entanglement entropy associated with the presence of the horizon~\cite{Sorkin:1986,Srednicki:1993,Solodukhin:2011,Speranza:2023} which is given by equation~\eqref{area entropy} and the second one to the von Neumann entropy of the matter.

The von Neumann entropy of matter fields is in general a complicated non-local expression. For the quantum fields with a fixed ultraviolet point it holds~\cite{Bisognano:1976,Jacobson:2015,Casini:2016,Speranza:2016,Arias:2016,Arzano:2020,Ribisi:2023,Chakraborty:2023}
\begin{equation}
\label{von Neumann ncft}
\delta S_{\text{vN}}=\frac{2\pi}{\hbar\hat{\kappa}}\int_{\Sigma_0}\delta\langle\hat{T}^{\mu\nu}\rangle\hat{\zeta}_{\mu}\hat{n}_{\nu}\text{d}^{D-1}\Sigma+\delta\hat{X}+O\left(l^{D+2}\right),
\end{equation}
where we recall that $\hat{\zeta}_{\mu}$ is the conformal Killing vector of the LCD. The term $\delta\hat{X}$ is a rather complicated, but explicitly known, spacetime scalar dependent on the LCD's size parameter $l$~\cite{Casini:2016,Speranza:2016}. For conformal field theories, $\delta\hat{X}$ identically vanishes.

To compute the perturbation of the horizon entanglement entropy, $\eta\delta\hat{\mathcal{A}}$, one must first specify the class of isentropic metric perturbations to which the equilibrium condition~\eqref{equilibrium} applies. It has been shown that defining a Euclidean canonical ensemble for an LCD relies on  fixing the spatial volume of $\Sigma_0$~\cite{Jacobson:2023,Jacobson:2023b}. Likewise, the first law of LCDs includes a contribution from the volume perturbation, that has the form of a work term in the standard first law of thermodynamics~\cite{Jacobson:2019,Alonso:2022}. From both perspectives, it can be expected that equation~\eqref{equilibrium} holds only when the spatial volume of the LCD is held fixed. This condition has been indeed confirmed in the literature~\cite{Jacobson:2015,Bueno:2017}. Thence, to find the isentropic $\eta\delta\hat{\mathcal{A}}$, we must be able to also compute the volume of $\Sigma_0$, including the curvature corrections. The light-cone cut construction of an LCD does not uniquely fix $\Sigma_0$. Thus, in this section, we consider a different generalization of a causal diamond to a curved spacetime LCD, a geodesic LCD~\cite{Jacobson:2015,Jacobson:2017,Wang:2019} (alternatively, one would have to, rather arbitrarily, specify $\Sigma_0$ within the light-cone cut framework).

To construct a geodesic LCD, we choose a regular spacetime point $P$ and a local direction of time $n^{\mu}$. Next, we send out geodesics of affine length $l$ in every direction orthogonal to $n^{\mu}$. Given that $l$ is much smaller than the curvature length scale, these geodesics do not intersect and form a spacelike $\left(D-1\right)$-dimensional geodesic ball $\Sigma_0$, whose boundary is an approximate $\left(D-2\right)$-sphere $\mathcal{B}$. The geodesic LCD then corresponds to the union of the past and future Cauchy developments of $\Sigma_0$ (its domain of dependence).

For a geodesic LCD we can easily evaluate the perturbations of the area and of the volume using expansion of the metric around $P$ in Riemann normal coordinates~\cite{Brewin:2009}
\begin{equation}
\label{RNC}
g_{\mu\nu}(x)=\eta_{\mu\nu}-\frac{1}{3}R_{\mu\alpha\nu\beta}\left(P\right)x^{\alpha}x^{\beta}+O\left(x^3\right),
\end{equation}
where $\eta_{\mu\nu}$ denotes the flat spacetime metric. The detailed computation has been carried out previously~\cite{Jacobson:2015} and, for the sake of brevity, we do not repeat it here. The final result for the area perturbation at fixed volume $\hat{\mathcal{V}}$ reads
\begin{equation}
\delta\hat{\mathcal{A}}\vert_{\hat{\mathcal{V}}}=-\frac{\Omega_{D-2}l^{D}}{D^2-1}\left(\hat{G}_{\mu\nu}\hat{n}^{\mu}\hat{n}^{\nu}-\lambda\right)+O\left(l^{D+2}\right),
\end{equation}
where $O\left(l^{D+2}\right)$ appears due to the subleading curvature-dependent corrections.

We now have everything we need to evaluate the equilibrium condition $\delta S_{\text{vN}}+\eta\delta\hat{\mathcal{A}}\vert_{\hat{\mathcal{V}}}=0$. It reads, up to $O\left(l^{D+2}\right)$ subleading terms,
\begin{equation}
\frac{2\pi}{\hbar}\left(\delta\langle\hat{T}_{\mu\nu}\rangle\hat{n}^{\mu}\hat{n}^{\nu}+\delta\hat{X}\right)-\eta\left(\hat{G}_{\mu\nu}\hat{n}^{\mu}\hat{n}^{\nu}-\lambda\right)=0,
\end{equation}
from which we obtain
\begin{equation}
\label{precursor-e}
8\pi G\left(\delta\langle\hat{T}_{\mu\nu}\rangle-\delta\hat{X}\hat{g}_{\mu\nu}\right)-\hat{G}_{\mu\nu}-\lambda\hat{g}_{\mu\nu}=0,
\end{equation}
where we again defined the Newton constant in terms of $\eta$, as $G=1/\left(4\hbar\eta\right)$, and the arbitrariness of the unit, timelike vector field $\hat{n}^{\mu}$ allowed us to remove the contractions with $\hat{n}^{\mu}$ (see Appendix \ref{n-dependence}). 
	
To complete the derivation, we need to determine the equilibrium value of the local curvature $\lambda$ which corresponds to an isentropic perturbation. To do so, we simply take the trace of equations~\eqref{precursor-e}, obtaining
\begin{equation}
\lambda=8\pi G\left(\frac{1}{D}\delta\langle\hat{T}\rangle-\delta\hat{X}\right)+\frac{D-2}{2D}\hat{R}, \label{lambda}
\end{equation}
i.e. the equilibrium value of $\lambda$ corresponds to a combination of the scalar curvature and terms determined by the perturbation of the matter fields. As an aside, for conformal matter fields satisfying the local energy conservation, we have \mbox{$\delta\langle\hat{T}\rangle=\delta\hat{X}=0$}, which leads to a much simpler expression for $\lambda$, namely \mbox{$\lambda=\left(D-2\right)\hat{R}/\left(2D\right)$}. Then, the scalar curvature of the unperturbed, locally maximally symmetric spacetime and the perturbed spacetime are equal and we have $\delta\hat{R}=0$.

Finally, plugging $\lambda$ into equations~\eqref{precursor-e} yields the traceless equations governing the gravitational dynamics
\begin{equation}
\hat{R}_{\mu\nu}-\frac{1}{D}\hat{R}=8\pi G\left(\delta\langle\hat{T}_{\mu\nu}\rangle-\frac{1}{D}\delta\langle\hat{T}\rangle\hat{g}_{\mu\nu}\right),
\end{equation}
valid at the point $P$. The strong equivalence principle guarantees that these equations hold throughout the spacetime.

A subtle issue should be noted. The curvature of the perturbed spacetime includes the contribution of $\lambda$. Equation~\eqref{lambda} connects $\lambda$ with $\delta\hat{X}$, which in general explicitly depends on the arbitrary size parameter $l$ of the LCD~\cite{Jacobson:2015,Speranza:2016,Casini:2016}. Then, the traceless Ricci tensor also depends on $l$ and this arbitrary parameter enters the equations governing the gravitational dynamics. This issue does not occur for conformally invariant matter fields which obey $\delta\hat{X}=0$.

All the arguments we gave in the previous subsection for the recovery of Weyl transverse gravity apply here as well. In particular, we again rely on the construction of a local causal diamond which is insensitive to the overall conformal factor of the metric, leading to the Weyl invariance of the resulting equations. Moreover, since the equations are traceless, they do not explicitly enforce the local energy conservation and the cosmological constant appears as an on shell integration constant, just like in Weyl transverse gravity. Lastly, the derivation is again insensitive to the shifts of entropy by a universal constant, just like in Weyl transverse gravity~\cite{Alonso:2022}.

Our conclusions are in fact consistent with ~\cite{Jacobson:2015} which introduced the entanglement equilibrium approach. It already explicitly shows the main feature of Weyl transverse gravity, namely, the appearance of $\Lambda$ as an integration constant. Moreover, while reference~\cite{Jacobson:2015} explicitly assumes Diff invariance, its results are actually independent of it. Rather, it relies on the strong equivalence principle, which is also incorporated in Weyl transverse gravity.

Notably, the equations governing the gravitational dynamics we derived have the quantum expectation value of the energy-momentum tensor as a source for the classical spacetime curvature. Therefore, we have in fact obtained the semiclassical traceless equations for Weyl transverse gravity.

\section{Equations of motion from Wald entropy in WTDiff-invariant gravity}
\label{WTDiff from TD}
	
In the previous section, we have argued that thermodynamics of LCDs encodes (semi)classical gravitational dynamics equivalent to Weyl transverse gravity. For Diff-invariant gravity it has been further shown that thermodynamics also encodes the equations of motion of any gravitational theory whose Lagrangian is a function of only the metric and the Riemann tensor~\cite{Bueno:2017,Svesko:2018,Svesko:2019}. The derivation of this result uses the Wald entropy prescription~\cite{Wald:1993,Wald:1994} corresponding to the given Lagrangian as an input. Then, one expects that the equations of motion for any WTDiff-invariant Lagrangian constructed from the auxiliary metric and Riemann tensor should also be obtainable in this way. Otherwise, the correspondence between thermodynamics and WTDiff symmetry we established in the previous section would be called into question, as it would be unable to reproduce all the results available in the Diff-invariant setup.
	
It is not immediately clear that the recovery of the equations of motion for modified WTDiff-invariant theories from their Wald entropy is possible. The first technical issue we face is that the original definition of Wald entropy in the covariant phase space approach applies only to local, Diff-invariant theories of gravity~\cite{Wald:1993,Wald:1994}. Thus, the first necessary step we already performed was to extend the covariant phase space construction and the definition of Wald entropy to arbitrary local, WTDiff-invariant theories of gravity in~\cite{Alonso:2022,Alonso:2022b}. Extending the covariant phase space formalism was necessary, but not sufficient by itself, since applying the resulting Wald entropy prescription to an LCD is not straightforward as we explain in the following. The conformal Killing vector $\zeta^{\mu}$~\eqref{conformal Killing} of an LCD clearly generates an infinitesimal diffeomorphism, which makes it a local symmetry generator for Diff-invariant gravity. However, $\zeta^{\mu}$ does not in general generate a transformation belonging to the WTDiff group\footnote{We might try to solve this issue by working with a Rindler wedge, which possesses a true (approximate) Killing vector belonging to the WTDiff group. However, the problems with applying thermodynamics of spacetime to Rindler wedges we mentioned in the previous section apparently make it impossible to derive equations of motion of modified gravitational theories from thermodynamics of Rindler wedges~\cite{Svesko:2018} (or, at least, such a derivation needs to be very fine-tuned and appears somewhat unnatural~\cite{Jacobson:2012}). Therefore, we need to work with LCDs.} (see, e.g.~\cite{Alvarez:2016,Alonso:2022} for longer discussions of this subtlety). One then has to be careful in applying the results of the covariant phase space construction to $\zeta^{\mu}$, since it does not correspond to a local symmetry of WTDiff-invariant gravity~\cite{Wald:1990,Alonso:2022,Alonso:2022b}. Nevertheless, $\zeta^{\mu}$ lies in the WTDiff group in a flat background (it generates a Weyl transformation of the dynamical metric $g_{\mu\nu}$). Using this observation, we showed that the WTDiff-invariant Wald entropy prescription indeed works for LCDs~\cite{Alonso:2022,Alonso:2022b}. Therefore, we can employ it to derive the equations governing the gravitational dynamics. As in the previous section, we perform the derivation by (the same) two different approaches: the physical process one and the entanglement equilibrium one.
	
Before proceeding, let us stress one subtle issue associated with deriving the equations governing gravitational dynamics from Diff-invariant Wald entropy. The cosmological constant term in a Diff-invariant gravitational Lagrangian does not affect Wald entropy. Consequently, the equations one derives from thermodynamics fail to reproduce this term in the Lagrangian. Instead, the cosmological constant arises as an on-shell integration constant in the process of solving the equations and has no connection with the fixed parameter in the Lagrangian that would enter the Euler-Lagrange equations. This failure to completely reconstruct the information contained in the Lagrangian of course disappears in WTDiff-invariant gravity. In this case, both the Euler-Lagrange equations and the thermodynamically derived equations are traceless and unaffected by any constant parameter in the Lagrangian. Therefore, as we will show, both procedures to derive the gravitational equations are fully consistent with each other.

\subsection{Physical process approach}
	
We first consider a physical process derivation. Our setup is essentially the same as in subsection~\ref{physical process Weyl transverse gravity}, i.e., we again study the change of entropy of a light-cone cut LCD between times $t=-\epsilon$ and $t=0$  (with $\epsilon\ll l$). Therefore, we consider a slice of the null boundary of the LCD sketched in figure~\ref{slice}. However, the Raychaudhuri equation approach we introduced in subsection~\ref{physical process Weyl transverse gravity} does not straightforwardly generalize to generic Wald entropy expressions~\cite{Eling:2006,Jacobson:2012}. Instead, we use the approach based on a conformal Killing identity~\cite{Svesko:2018,Svesko:2019}, which we adapt for the WTDiff-invariant setup.
	
The equations of motion are encoded in the equilibrium condition applied to the changes of the matter entropy and the Wald entropy of the horizon. We have already computed the change in matter Clausius entropy in subsection~\ref{physical process Weyl transverse gravity}, obtaining equation~\eqref{dSC}.
	
Wald entropy for the class of theories described by the action~\eqref{I WTDiff} obeys (see ~\cite{Alonso:2022b} for a discussion of the subtleties involved in the defining it)
\begin{align}
S_{\text{W}}\left(t\right)&=\frac{2\pi}{\hbar\kappa}\int_{\mathcal{B}_{t}}Q^{\nu\mu}_{\zeta}\text{d}\tilde{B}_{\mu\nu}, \label{S Wald} \\
Q^{\nu\mu}_{\zeta}&=2\tilde{E}^{\nu\mu\rho}_{\quad\;\:\sigma}\tilde{\nabla}_{\rho}\zeta^{\sigma}-4\tilde{\nabla}_{\rho}\tilde{E}^{\nu\mu\rho}_{\quad\;\:\sigma}\zeta^{\sigma},
\end{align}
where $Q^{\nu\mu}_{\zeta}$ denotes the (background-dependent) Noether charge corresponding to the conformal Killing vector $\zeta^{\mu}$, we define the oriented area element \mbox{$\text{d}\tilde{B}_{\mu\nu}=\left(\tilde{n}_{\mu}\tilde{m}_{\nu}-\tilde{n}_{\nu}\tilde{m}_{\mu}\right)\text{d}^{D-2}\tilde{\mathcal{A}}$}, where $\tilde{m}^{\mu}$ is the WTDiff-invariant unit spatial normal to $\mathcal{B}$; finally, the tensor $\tilde{E}_{\mu}^{\;\:\nu\rho\sigma}=\partial L/\partial\tilde{R}^{\mu}_{\;\:\nu\rho\sigma}$ has been introduced in equation~\eqref{dL/dR}. It is easy to see that the entropy of an LCD is time-dependent (its area expands). Then, even though $\zeta^{\mu}$ vanishes at the bifurcation surface $t=0$ of the LCD's boundary (see equation~\eqref{conformal Killing}), the term proportional to $\zeta^{\mu}$ does contribute to entropy. As an aside, the terms of this form are also crucial for the recent proposal for entropy of dynamical black holes~\cite{Hollands:2024}. By applying the generalized Stokes theorem, we obtain for the change of Wald entropy between the spatial spheres $\mathcal{B}_{-\epsilon}$ and $\mathcal{B}$
\begin{equation}
\label{DSW}
\Delta S_{\text{W}}=\frac{2\pi}{\kappa}\int_{-\epsilon}^{0}\text{d}t\int_{\mathcal{B}_{t}}\text{d}^{D-2}\tilde{\mathcal{A}}\tilde{k}_{-}^{\mu}\tilde{\nabla}_{\nu}Q^{\nu\mu}_{\zeta},
\end{equation}
where $\tilde{k}_{-}^{\mu}$ is a future-pointing, WTDiff-invariant null normal to the horizon. 

Our strategy is to rewrite $\Delta S_{\text{W}}$ in a form that does not include derivatives of $\zeta^{\mu}$ (present in $\tilde{\nabla}_{\nu}Q^{\nu\mu}_{\zeta}$), which will allow us to straightforwardly carry out the integration. To remove the derivatives of $\zeta^{\mu}$, we invoke the (approximate) conformal Killing identity~\cite{Svesko:2019}
\begin{align}
\nonumber \tilde{\nabla}_{\nu}\tilde{\nabla}_{\rho}\zeta_{\sigma}=&~\tilde{R}_{\lambda\nu\rho\sigma}\zeta^{\lambda}+\frac{1}{D}\tilde{g}_{\rho\sigma}\tilde{\nabla}_{\nu}\tilde{\nabla}_{\lambda}\zeta^{\lambda}+\frac{1}{D}\tilde{g}_{\nu\sigma}\tilde{\nabla}_{\rho}\tilde{\nabla}_{\lambda}\zeta^{\lambda} \\
&-\frac{1}{D}\tilde{g}_{\nu\rho}\tilde{\nabla}_{\sigma}\tilde{\nabla}_{\lambda}\zeta^{\lambda} + O(\epsilon^2l^D). \label{Killing identity}
\end{align}
Applying this identity to equation~\eqref{DSW} yields
\begin{align}
\nonumber \Delta S_{\text{W}}=&~\frac{4\pi}{\hbar\kappa}\int_{-\epsilon}^{0}\text{d}t\int_{\mathcal{B}_{t}}\text{d}^{D-2}\tilde{\mathcal{A}}\tilde{k}_{-}^{\mu}
\\
\nonumber &\times\bigg[-\tilde{E}_{\nu}^{\;\:\lambda\rho\sigma}\tilde{R}_{\mu\lambda\rho\sigma}\zeta^{\nu}+2\tilde{\nabla}_{\rho}\tilde{\nabla}_{\sigma}\tilde{E}_{\mu\;\;\;\nu}^{\;\:\rho\sigma}\zeta^{\nu} \\
&+\frac{64\pi G}{D}\tilde{\nabla}_{\rho}\tilde{\nabla}_{\lambda}\zeta^{\lambda}\tilde{E}_{\mu\nu}^{\;\:\;\:\rho\nu}\bigg], \label{Wald flux}
\end{align}
where we discarded the $O\left(\epsilon^3\right)$ terms. In general, other contributions would appear due to the fact that $\zeta^{\mu}$ is only an approximate conformal Killing vector and does not precisely satisfy the conformal Killing identity~\eqref{Killing identity}. However, these extra terms can be removed by adding suitable curvature-dependent terms (which disappear in flat spacetime) to the definition of $\zeta^{\mu}$~\eqref{conformal Killing}. The procedure has been worked out in detail in the Diff-invariant setup and translates without any changes to the WTDiff-invariant case we consider~\cite{Svesko:2019}.
	
We specialize to gravitational theories whose Lagrangian contains a term of the form $\tilde{R}\omega/\left(16\pi G\right)$, i.e., on those that directly generalize Weyl transverse gravity. For any such theory, the last term on the right hand side of equation~\eqref{Wald flux} is the only one that does not vanish in flat spacetime (for Lagrangians without a term linear in curvature, all three terms vanish and the following discussion does not apply). Therefore, much like in the special case of Weyl transverse gravity (see equation~\eqref{area expansion} and the accompanying discussion) we split the change of the Wald entropy in a flat spacetime contribution and a contribution induced by the spacetime curvature
\begin{align}
\Delta S_{\text{W}}=&\Delta S_{\text{flat}}+\Delta S_{\text{curv}}, \\
\Delta S_{\text{flat}}=&\frac{4\pi}{\hbar\kappa}\frac{4}{D}\int_{-\epsilon}^{0}\text{d}t\int_{\mathcal{B}_{t}}\text{d}^{D-2}\tilde{\mathcal{A}}\tilde{k}_{-}^{\mu}\tilde{\nabla}_{\rho}\tilde{\nabla}_{\lambda}\zeta^{\lambda}\tilde{E}_{\mu \nu}^{\;\;\;\:\rho \nu}, \label{S irr} \\
\nonumber \Delta S_{\text{curv}}=&\frac{1}{4\hbar G\kappa}\int_{-\epsilon}^{0}\text{d}t\int_{\mathcal{B}_{t}}\text{d}^{D-2}\tilde{\mathcal{A}}\tilde{k}_{-}^{\mu}\big[\tilde{E}_{\nu}^{\;\;\lambda\rho\sigma}\tilde{R}_{\mu\lambda\rho\sigma} \\
&-2\tilde{\nabla}_{\rho}\tilde{\nabla}_{\sigma}\tilde{E}_{\mu\;\;\;\nu}^{\;\;\rho\sigma}\big]\zeta^{\nu}. \label{S rev}
\end{align}
In the following, as in subsection~\ref{WTG from TD}, we disregard the flat spacetime term $\Delta S_{\text{flat}}$ (for its possible interpretation, see~\cite{Svesko:2019}). In any case, only $\Delta S_{\text{curv}}$ is connected with a matter Clausius entropy flux across the horizon. We thus study the thermodynamic equilibrium condition
\begin{equation}
\label{equilibrium 2}
\Delta S_{\text{C}}+\Delta S_{\text{curv}}=0.
\end{equation}

The integrals in~\eqref{S rev} can be performed straightforwardly to yield
\begin{align}
\nonumber \Delta S_{\text{curv}}=&-\epsilon^2\frac{D\Omega_{D-2}}{8\hbar G\left(D-1\right)}l^{D-2}\left(\tilde{H}_{\mu\nu}-\frac{1}{D}\tilde{H}\tilde{g}_{\mu\nu}\right)\tilde{n}^{\mu}\tilde{n}^{\nu} \\
&+O\left(\epsilon^2l^D\right), \label{DSC}
\end{align}
where the $O\left(\epsilon^2l^D\right)$ terms come from the higher order contributions in the Riemann normal coordinate expansion and we identified the symmetric tensor $\tilde{H}_{\mu\nu}$ defined by equation~\eqref{H}.

Next, we plug this result and the expression~\eqref{dSC} for $\Delta S_{\text{C}}$ into the equilibrium condition~\eqref{equilibrium 2}. Since, as before, $\tilde{n}^{\mu}$ is an arbitrary, timelike, unit, WTDiff-invariant vector field, we remove the contractions with it. Lastly, the Einstein equivalence principle guarantees that the resulting equations for gravitational dynamics valid in $P$ hold in every regular spacetime point\footnote{The strong equivalence principle does not apply to modified theories of gravity~\cite{Casola:2014,Casola:2015,equivalence}. In the thermodynamic context, it shows up in the position-dependent density of Wald entropy.}, eliminating the dependence of the quantities on $P$. In total, we arrive at the following traceless equations valid throughout the spacetime
\begin{equation}
\tilde{H}_{\mu\nu}-\frac{1}{D}\tilde{H}\tilde{g}_{\mu\nu}=8\pi G\left(\tilde{T}_{\mu\nu}-\frac{1}{D}\tilde{T}\tilde{g}_{\mu\nu}\right). \label{phys equations}
\end{equation}
We can see that we have reproduced the traceless equations of motion~\eqref{traceless WTDiff} of a local, WTDiff-invariant theory of gravity whose Lagrangian is an arbitrary function of $\tilde{g}_{\mu\nu}$ and $\tilde{R}^{\mu}_{\;\:\nu\rho\sigma}$. We stress that both the thermodynamically derived equations~\eqref{phys equations} and the Euler-Lagrange equations~\eqref{traceless WTDiff} are traceless and recover the cosmological constant as an on-shell integration constant. Thence, unlike in the Diff-invariant case, Wald entropy suffices to recover all the information contained in the WTDiff-invariant gravitational Lagrangian~\eqref{I WTDiff}.

\subsection{Entanglement equilibrium approach}
	
The entanglement equilibrium derivation we analyzed in subsection~\ref{entanglement equilibrium} can also be generalized to local, WTDiff-invariant theories of gravity given by the Lagrangian~\eqref{I WTDiff}. Our treatment largely follows the method developed for Diff-invariant gravity~\cite{Bueno:2017}, which we modify for the WTDiff-invariant setup.
	
The renormalized entanglement entropy associated with the horizon of a geodesic LCD takes the same form as the Wald entropy of certain modified gravity theories~\cite{Bousso:2016}. Therefore, we have the following entanglement equilibrium condition
\begin{equation}
\label{equilibrium Wald}
\delta S_{\text{W}}+\delta S_{\text{vN}}=0,
\end{equation}
where $\delta S_{\text{W}}$ denotes the Wald entropy perturbation, and the matter von Neumann entropy perturbation $\delta S_{\text{vN}}$ obeys equation~\eqref{von Neumann ncft}. As we discussed in subsection~\ref{entanglement equilibrium}, the equilibrium state of the LCD corresponds to a locally maximally symmetric spacetime with curvature \mbox{$\hat{G}^{\text{MSS}}_{\mu\nu}=-\lambda\hat{g}^{\text{MSS}}_{\mu\nu}$}, where $\lambda$ in principle depends on the position and size of the LCD as before.

We now need to evaluate the perturbation of Wald entropy for the bifurcate $\left(n-2\right)$-surface $\mathcal{B}$  of the horizon. Then, plugging the result into the equilibrium condition~\eqref{equilibrium Wald} will allow us to obtain the equations of motion.
	
Wald entropy is given by equation~\eqref{S Wald}. Since $\zeta^{\mu}$ vanishes on the bifurcate surface $\mathcal{B}$, a generic perturbation of equation~\eqref{S Wald} on this surface yields
\begin{equation}
\delta S_{\text{W}}=\frac{4\pi}{\hbar\kappa}\delta\int_{\mathcal{B}}\tilde{E}^{\nu\mu\rho}_{\quad\;\:\sigma}\tilde{\nabla}_{\rho}\zeta^{\sigma}\text{d}\mathcal{B}_{\mu\nu}.
\end{equation}
Since $l$ is much smaller than the local curvature length scale, we can expand $\tilde{E}^{\nu\mu\rho}_{\quad\;\:\sigma}$ in powers of $l$ around the LCD's center $P$, keeping only the first three terms in the expansion
\begin{align}
\nonumber \delta S_{\text{W}}=&~\frac{4\pi}{\hbar\kappa}\delta 
\int_{\mathcal{B}}\bigg(\tilde{E}^{\nu\mu\rho}_{\quad\;\:\sigma}+l\tilde{m}^{\lambda}\tilde{\nabla}_{\lambda}\tilde{E}^{\nu\mu\rho}_{\quad\;\:\sigma} \\
&+\frac{1}{2}l^2\tilde{m}^{\lambda}\tilde{m}^{\tau}\tilde{\nabla}_{\lambda}\tilde{\nabla}_{\tau}\tilde{E}^{\nu\mu\rho}_{\quad\;\:\sigma}\bigg)\tilde{\nabla}_{\rho}\zeta^{\sigma}\text{d}\mathcal{B}_{\mu\nu}, \label{entropy variation}
\end{align}
where we evaluate all the tensors at $P$. For computational convenience, we split $\tilde E^{\;\:\nu\rho\sigma}_{\mu}$ into the part corresponding to Weyl transverse gravity and the higher order corrections we denote by $\tilde{F}^{\;\:\nu\rho\sigma}_{\mu}$, i.e.,
\begin{equation}
\label{define F}
\tilde{E}^{\;\:\nu\rho\sigma}_{\mu}=\frac{1}{32\pi G}\left(\delta^{\rho}_{\mu}\tilde{g}^{\nu\sigma}-\delta^{\sigma}_{\mu}\tilde{g}^{\nu\rho}+2\tilde{F}^{\;\:\nu\rho\sigma}_{\mu}\right).
\end{equation}
Note that, since $\tilde{F}^{\;\:\nu\rho\sigma}_{\mu}$ vanishes in a maximally symmetric spacetime $\delta \tilde{F}^{\;\:\nu\rho\sigma}_{\mu}= \tilde{F}^{\;\:\nu\rho\sigma}_{\mu}$, so we obtain
\begin{align}
\nonumber \delta S_{\text{W}}=&~\frac{\delta\tilde{\mathcal{A}}_{\mathcal{B}}}{4\hbar G}-\frac{\Omega_{D-2}l^{D-2}}{4\hbar G\left(D-1\right)}\tilde{n}^{\mu}\tilde{n}^{\nu} \\
&\times\bigg[\tilde{W}_{\mu\nu}-\frac{l^2}{\left(D+1\right)}\tilde{\nabla}_{\rho}\tilde{\nabla}_{\sigma}\tilde{F}^{\;\:\rho\sigma}_{\mu\;\;\;\:\nu}\bigg], \label{dS W}
\end{align}
where $\tilde{F}^{\;\:\nu\rho\sigma}_{\mu}$ now corresponds to its value for the perturbed metric. To simplify the notation, we introduced a tensor
\begin{align}
\tilde{W}_{\mu\nu}=&~\left(1+\frac{l^2}{2\left(D+1\right)}\tilde{h}^{\rho \sigma} \tilde{\nabla}_\rho \tilde{\nabla}_\sigma \right)\tilde{F}^{\lambda}_{\;\:\mu\lambda\nu}, \label{W tensor}
\end{align}
whose significance will become clear in the following.

In the special case of Weyl transverse gravity, we have seen that the equilibrium condition~\eqref{equilibrium Wald} only applies to perturbations that hold fixed the WTDiff-invariant volume of the geodesic $\left(D-1\right)$-dimensional ball $\Sigma_0$ (as its perturbation corresponds to a work term in the first law of causal diamonds~\cite{Jacobson:2019,Jacobson:2019b,Alonso:2022,Alonso:2022b}). It has been shown that the former condition translates directly to the case of more general WTDiff-invariant theories of gravity~\cite{Bueno:2017}. However, rather than the spatial volume of $\Sigma_0$, one needs to keep fixed its generalized volume $\tilde{\mathcal{W}}$, whose perturbation equals~\cite{Bueno:2017,Alonso:2022b}
\begin{equation}
\label{generalised volume}
\delta\tilde{\mathcal{W}}_{\Sigma_0}=\delta\tilde{\mathcal{V}}_{\Sigma_0}+\frac{\Omega_{D-2}l^{D-1}}{4\left(D-1\right)\left(D-2\right)}\tilde{W}_{\mu\nu}\tilde{n}^{\mu}\tilde{n}^{\nu}=0,
\end{equation}
where $\tilde{\mathcal{V}}_{\Sigma_0}$ denotes the geometric, WTDiff-invariant volume. For Weyl transverse gravity, $\tilde{W}_{\mu\nu}$ vanishes and the generalized volume reduces to the geometric one. The generalized volume appears in the first law of LCDs in modified theories of gravity instead of the standard volume~\cite{Bueno:2017}. It also must be held fixed to define an LCD Euclidean canonical ensemble in modified theories of gravity~\cite{Tavlayan:2023}. Thence, it plays the same role for modified theories of gravity as the geometric WTDiff-invariant volume does for Weyl transverse gravity. In conclusion, the entanglement equilibrium condition~\eqref{equilibrium Wald} applies to perturbations obeying $\delta\tilde{\mathcal{W}}=0$.

The Wald entropy perturbation at fixed generalized volume, $\delta\tilde{\mathcal{W}}=0$, has been show to equal~\cite{Bueno:2017}
\begin{align}
\nonumber \delta S_{\text{W}}\big\vert_{\delta\tilde{\mathcal{W}}=0}=&\frac{\Omega_{D-2}l^{D}}{4\hbar G\left(D^2-1\right)}
\tilde{n}^{\mu}\tilde{n}^{\nu}\\
&\times\Big[\tilde{G}_{\mu\nu}-\lambda\tilde{g}_{\mu\nu} +2\tilde{\nabla}_{\rho}\tilde{\nabla}_{\sigma}\tilde{F}^{\;\:\rho\sigma}_{\mu\;\;\;\:\nu}\Big].
\end{align}
	
At this point, we have all the ingredients to evaluate the entanglement equilibrium condition~\eqref{equilibrium Wald}. After some straightforward simplifications, it reads
\begin{align}
\nonumber \Big[\tilde{G}_{\mu\nu}-\lambda\tilde{g}_{\mu\nu}+2\tilde{\nabla}_{\rho}\tilde{\nabla}_{\sigma}\tilde{F}^{\;\:\rho\sigma}_{\mu\;\;\;\:\nu}& \\
-8\pi G\left(\delta\langle\tilde{T}_{\mu\nu}\rangle-\delta\tilde{X}\right)\Big]\tilde{n}^{\mu}\tilde{n}^{\nu}&=0. \label{equilibrium Wald 2}
\end{align}
We can once again remove the contractions with an arbitrary, unit timelike vector field $\tilde{n}^{\mu}$ (see Appendix~\ref{n-dependence}). Taking the trace of the equations determines the equilibrium value of $\lambda$ corresponding to an isentropic perturbation
\begin{equation}
\lambda=8\pi G\left(\frac{1}{D}\delta\langle\tilde{T}\rangle-\delta\tilde{X}\right)+\frac{D-2}{2D}\tilde{R}+\frac{2}{D}\tilde{\nabla}_{\rho}\tilde{\nabla}_{\sigma} \tilde{F}^{\;\:\rho\lambda\sigma}_{\lambda}.
\end{equation}
Plugging $\lambda$ back into the entanglement equilibrium condition~\eqref{equilibrium Wald 2}, we obtain the traceless equations
\begin{align}
\tilde{H}_{\mu\nu}^{(1)}-\frac{1}{D}\tilde{H}^{(1)}\tilde{g}_{\mu\nu}=8\pi G\left(\delta\langle\tilde{T}_{\mu\nu}\rangle-\frac{1}{D}\delta\langle\tilde{T}\rangle\tilde{g}_{\mu\nu}\right).
\end{align}
where we defined $\tilde{H}_{\mu\nu}^{(1)}$ as the part of the symmetric tensor $\tilde{H}_{\mu\nu}$ linear in the Riemann tensor, i.e.,
\begin{equation}
\tilde{H}_{\mu\nu}^{(1)}=\tilde{R}_{\mu\nu}+2\tilde{\nabla}_{\rho}\tilde{\nabla}_{\sigma}\tilde{F}^{\;\;\:\rho\sigma}_{(\mu\;\;\;\:\nu)}.
\end{equation}
We used the fact that tensor $\tilde{F}^{\;\:\rho\sigma}_{\mu\;\;\;\:\nu}$ is itself at least linear in the Riemann tensor (as can be seen from equation~\eqref{dL/dR} and the definition of $\tilde{F}^{\;\:\rho\sigma}_{\mu\;\;\;\:\nu}$~\eqref{define F}). Therefore, we discarded any contractions of $\tilde{F}^{\;\:\rho\sigma}_{\mu\;\;\;\:\nu}$ with the Riemann tensor. The Einstein equivalence principle guarantees that these equations hold throughout the spacetime. We have thus recovered the linearized, semiclassical traceless equations of motion for local, WTDiff-invariant theory of gravity whose Lagrangian is an arbitrary function of $\tilde{g}_{\mu\nu}$ and $\tilde{R}^{\mu}_{\;\:\nu\rho\sigma}$. In total, we have shown that the WTDiff-invariant thermodynamics of spacetime is fully consistent both in the physical process and in the equilibrium approach.

\subsection{Comparison of both approaches}
\label{comparison}

To conclude this section, we briefly address the (in)equivalence of the physical process and the entanglement equilibrium approaches to deriving the equations governing gravitational dynamics. While both approaches recover the gravitational dynamics from equilibrium conditions applied to LCDs, they differ in two key aspects. First, the physical process approach evaluates the equilibrium conditions for a slice of the null boundary of the LCD, the entanglement equilibrium approach does so for a spacelike ball. Second, the former approach works in a generic curved spacetime, whereas the latter starts in a (locally) maximally symmetric spacetime and introduces a small perturbation of it.

In section~\ref{WTG from TD}, we have seen that both approaches equivalently recover the equations of motion of Weyl transverse gravity (although the resulting equations are semiclassical only for the entanglement equilibrium approach). However, a difference occurs for modified theories of gravity we studied in this section. The entanglement equilibrium approach allows us to derive only the linearized equations of motion~\cite{Bueno:2017}, whereas the physical process approach recovers the full non-linear dynamics~\cite{Svesko:2018,Svesko:2019}. This outcome is not so surprising, since the entanglement equilibrium approach linearizes the equilibrium conditions in a small perturbation away from the locally maximally symmetric spacetime. Therefore, we can conclude that the physical process approach and the entanglement equilibrium approach are fundamentally distinct and in general yield different results. The reason both approaches equivalently recover the full non-linear dynamics of Weyl transverse gravity is likely related to the strong equivalence principle, which severely constrains the possible gravitational dynamics (as we argued, the only two possibilities are general relativity and Weyl transverse gravity). Then it becomes possible to infer the non-linear dynamics of Weyl transverse gravity even from the entanglement equilibrium approach.

\section{Discussion}
\label{discussion}
	
In this work, we have constructed a framework of local causal structures with the minimal necessary assumptions on the geometry and thermodynamic properties to show that equilibrium conditions imposed on LCDs encode WTDiff-invariant gravitational dynamics. We have performed the derivation in two independent ways, first using a physical process approach and then an entanglement equilibrium approach. In contrast to previous works on the subject, our derivation does not involve any \textit{a priori} assumptions about the local symmetries of gravity. Furthermore, we have verified that Weyl transverse gravity is in turn consistent with the two assumptions necessary to derive the gravitational dynamics from thermodynamic tools. Indeed, we previously derived explicitly the Wald entropy corresponding to the horizon of an LCD, showing it to be indeed proportional to its (WTDiff-invariant) area~\cite{Alonso:2022}. Moreover, this argument is completed with the discussion carried out in a related work~\cite{equivalence}, where we show how Weyl transverse gravity incorporates the strong equivalence principle.

Our case for WTDiff invariance can then be considered fully self-consistent and complete. The only bit that might appear to be missing is the explicit emergence of Weyl symmetry from thermodynamics. However, saying anything about the behavior of entropy of local causal horizons under Weyl transformations already involves a conscious choice about the behavior of the metric $\hat{g}_{\mu\nu}$ (that can a priori be either Diff- or WTDiff-invariant), since entropy is proportional to the area $\hat{\mathcal{A}}$ measured with respect to this metric. Therefore, the indirect arguments for Weyl invariance we offer probably cannot be further improved within the framework we use, although we we think that they are compelling enough on their own.

As an aside, the entanglement equilibrium derivation also allowed us to obtain the semiclassical equations for Weyl transverse gravity.

Independently of this main result, we have further derived the equations of motion for any WTDiff-invariant Lagrangian such that $L=L\left(\tilde{g}^{\mu\nu},\tilde{R}^{\mu}_{\;\:\nu\rho\sigma}\right)$ from the corresponding Wald entropy. While our approach builds on the methods previously developed in the Diff-invariant setup, it has one important advantage. For the Diff-invariant case, by deriving equations of motion from Wald entropy one fails to recover the cosmological constant, which is a fixed constant parameter in the Lagrangian. In the WTDiff-invariant setup, the Lagrangian carries no information about the cosmological constant (it is a global degree of freedom), and we can recover the full equations of motion from Wald entropy.

This work should not be understood as stating that the gravitational dynamics must be WTDiff invariant. It is equally possible that the local equilibrium conditions (together with the equivalence principle) simply do not contain enough information to fully recover the dynamics. Specifically, to obtain Diff-invariant gravitational dynamics from the local equilibrium conditions, one would need to introduce two additional requirements; the local energy-momentum conservation and a fixed value of the cosmological constant. In any case, given that Weyl transverse gravity has originally appeared in the context of field theoretical approach to gravity~\cite{Alvarez:2006} and as a possible resolution of some of the problems related to the cosmological constant~\cite{Unruh:1989,Finkelstein:2001,Barcelo:2014,Carballo:2015,Barcelo:2018} (since vacuum energy does not gravitate), it is remarkable that it also naturally emerges from thermodynamics of local causal horizons.
	
The physical process derivation of gravitational dynamics we introduced also allows for further generalizations. First, we have set the initial shear and expansion of the horizon to zero. However, it has been proposed that the equations for the gravitational dynamics can be derived even with the shear terms included, as they simply lead to further irreversible entropy production~\cite{Chirco:2010}. Furthermore, the recent proposal for dynamical black hole entropy suggests a way to deal with the expansion terms as well~\cite{Hollands:2024}. We intend to address both issues in a future work. Second, the change of the area of the light-cone cut LCD due to a metric perturbation in vacuum is proportional to the Bell-Robinson tensor, which has been proposed as a quasilocal measure of energy of the gravitational field~\cite{Wang:2019}. Then, using our physical process approach, it should be possible to derive modified vacuum equations for gravitational dynamics that relate the Einstein tensor to the Bell-Robinson tensor. We will report on this project in an upcoming paper.

\section*{Acknowledgements}

AA-S is funded by the Deutsche Forschungsgemeinschaft (DFG, German Research Foundation) — Project ID 51673086. ML is supported by the DIAS Post-Doctoral Scholarship in Theoretical Physics 2024 and by the Charles University Grant Agency project No. GAUK 90123. AA-S and LJG acknowledge support through Grants No. PID2020–118159 GB-C44 and PID2023-149018NB-C44 (funded by MCIN/AEI/10.13039/501100011033). LJG also acknowledges the support of the Natural Sciences and Engineering Research Council of Canada (NSERC).

\appendix

\section{Entropy of local causal diamonds}
\label{lcd entropy}

In this appendix, we review the arguments for assigning entropy to local causal horizons and, in particular, LCDs. A finite lifetime observer whose existence starts in the past apex $A_{\text{p}}$ of the LCD and ends in its future apex $A_{\text{f}}$ perceives the null boundary of the LCD as a causal horizon. Thence, it should be possible to assign entropy to this observer, quantifying their lack of information about the exterior region. Several approaches to computing entropy of a horizon indeed support the idea that local causal horizons (associated with a class of observers that perceive them) possess finite entropy~\cite{Sorkin:1986,Srednicki:1993,Jacobson:2003,Solodukhin:2011,Chakraborty:2016,Jacobson:2019,Jacobson:2023,Jacobson:2023b}.

In particular, there exists an established way to introduce entropy of a local causal horizon independently of the gravitational action. The Reeh-Schlieder theorem~\cite{Reeh:1961}  for quantum field theory in flat spacetime implies the existence of vacuum quantum entanglement between the interior and the exterior region of the LCD. Consequently, an observer restricted to the interior of the LCD measures a non-zero entanglement entropy~\cite{Sorkin:1986,Srednicki:1993,Solodukhin:2011}. Detailed calculations~\cite{Sorkin:1986,Srednicki:1993} show that this entropy diverges unless one introduces a suitable ultraviolet cutoff corresponding to a length scale $\varepsilon$. Upon introducing a cutoff, the entanglement entropy becomes finite and proportional to the horizon area $\mathcal{A}$, i.e., $S_{\text{e}}=\eta\mathcal{A}$~\cite{Sorkin:1986,Srednicki:1993,Solodukhin:2011}. The proportionality constant $\eta$ scales with the inverse square of the cutoff length, $\eta\propto1/\varepsilon^2$, and its value further depends on the matter fields present in the spacetime. If we take $\varepsilon$ to be of the order of the Planck length, $l_{\text{P}}$, the entanglement entropy of any causal horizon (including that of an LCD) becomes comparable with Bekenstein entropy, $S_{\text{B}}=\mathcal{A}/\left(4\hbar G\right)$. For this reason, quantum entanglement has also been suggested as a possible microscopic explanation of black hole entropy~\cite{Sorkin:1986,Srednicki:1993}. An important feature of entanglement entropy is that it has (to leading order) the same areal density for any boundary~\cite{Solodukhin:2011}. This outcome agrees with the results one obtains from the standard (gravitational action-dependent) approaches to computing entropy associated with a horizon~\cite{Gibbons:1977,Wald:1993,Jacobson:2023b}. However, some criticisms to the entanglement interpretation of horizon entropy has been put forward~\cite{Liberati:1997,Solodukhin:2011,Agullo:2023}:
\begin{itemize}
\item Entanglement entropy depends on which quantum fields are present in the spacetime. This criticism can be addressed in approaches that make the cutoff $\varepsilon$ (or the Planck length) also sensitive to the matter content of the theory, making the entanglement entropy independent of it~\cite{Susskind:1994,Jacobson:1994,Solodukhin:2011}.
\item The choice of Planck length as the ultraviolet cutoff can be motivated~\cite{Mead:1964,Garay:1994en,Hossenfelder:2013}, but it lacks a clear justification. Furthermore, the cutoff breaks the local Lorentz invariance of the theory. However, the calculation has also been rephrased using a covariant Pauli-Villars regulator, confirming the previously obtained cutoff-dependent results~\cite{Susskind:1994}.
\item It has been argued that, if the entanglement entropy explains the leading order term in black hole entropy, the vacuum fluctuations also significantly change black hole energy, breaking the self-consistency of the approach~\cite{Hooft:1985,Belgiorno:1996,Liberati:1997}. However, arguments against this viewpoint have been presented as well~\cite{Susskind:1994,Jacobson:1994,Demers:1995,Banks:2024}.
\item Many approaches to quantum gravity introduce some discretization of the spacetime, which only allows a finite subregion of it to have finitely many degrees of freedom. However, the Reeh-Schlieder theorem, which provides the theoretical justification of quantum entanglement between arbitrary spacelike separated subregions, only works for systems with infinitely many degrees of freedom. For systems with finitely many degrees of freedoms, it appears that quantum entanglement does not generically occur~\cite{Agullo:2023}. This observation undermines the entanglement interpretation of Bekenstein entropy assuming that spacetime is discretized. We are not aware of any way to refute this objection.
\end{itemize}

Although the entanglement interpretation of horizon entropy is often invoked in derivations of gravitational dynamics from thermodynamics, the derivation does not depend on the entropy interpretation in any way. All that one really needs to assume is the following. The observers perceiving a local causal horizon cannot access its exterior and should measure some entropy quantifying this fact. It should be possible to express this entropy in terms of the properties of the boundary, as it represents the only feature of the exterior accessible to the interior observer. Then, following the logic of the original proposal for black hole entropy~\cite{Bekenstein:1973}, we find entropy proportional to the horizon area to the leading order, $S=\eta\mathcal{A}$, to be the simplest possibility. Other terms proportional, e.g. to the extrinsic curvature of the boundary or its Euler characteristic can be present in principle~\cite{Solodukhin:2011,Dong:2014,Wall:2015,Hollands:2024}. However, for dimensional reasons, these terms scale either with higher powers of the size parameter of the horizon (e.g., with the spatial volume of the LCD), or with higher powers of the Planck length. Since, throughout this work, we focus on sufficiently small causal horizons and we neglect any quantum gravitational effects suppressed by powers of the Planck length, we can safely neglect any such term and keep the entropy proportional to the area.

We do not have to make any assumptions about the microscopic origin of this entropy. Moreover, we do not need to fix the proportionality constant $\eta$. The strong equivalence principle guarantees that $\eta$ is a universal constant. If that were not the case, one could devise a local experiment measuring the entropy density and obtain different results in two distinct locations, falsifying the statement of the principle. Then, rather than fixing $\eta$ to a specific value, we may instead \textit{define} the Newton gravitational constant in terms of $\eta$~\cite{Jacobson:2015}. Since we are deriving the gravitational dynamics from thermodynamics and not the other way around, we find this approach sensible regardless of whether gravity is a fundamental interaction or not.

As an aside, if one specifies the gravitational dynamics \textit{a priori}, a number of standard approaches show that LCDs indeed possess entropy proportional to its area to the leading order:
\begin{itemize}
\item Wald entropy density~\cite{Wald:1993,Wald:1994} has the same form for both a black hole Killing horizon and for a conformal Killing horizon of an LCD (both for Diff-invariant~\cite{Jacobson:2019} and for WTDiff-invariant~\cite{Alonso:2022,Alonso:2022b} theories of gravity). The entropy prescription follows from evaluating the Hamiltonian corresponding to evolution along the conformal Killing vector $\zeta^{\mu}$ for the interior of the LCD at $t=0$ (the spatial ball $\Sigma_0$).
\item Entropy of an LCD can be computed via the Cardy formula~\cite{Cardy:1986} as the symmetries of a wide class of null surfaces (including black hole horizons and local causal horizons) form a Virasoro algebra with a central charge, that is identical to the algebra of symmetries of a $2$-dimensional conformal field theory~\cite{Carlip:1999,Chakraborty:2016}. The Cardy formula valid for such a theory then allows us to compute the entropy from the central charge. It again yields the same entropy prescription for any causal horizon.
\item A Euclidean canonical ensemble has been constructed for LCDs~\cite{Jacobson:2023,Jacobson:2023b}. The method obtains the canonical partition function as a Euclidean path integral of the gravitational action in flat spacetime under the assumption of fixed volume of the spatial ball $\Sigma_0$ (implemented via a Lagrange multiplier). The resulting expression for entropy has again the same form as for a black hole.
\end{itemize}
Naturally, all the entropy calculations we have just listed rely on the knowledge of the gravitational action. Then, while they serve as supporting arguments for assigning entropy to LCDs, invoking them to derive the gravitational dynamics from thermodynamics clearly leads to a circular argument. We stress that the derivations we present in section~\ref{WTG from TD} do not in any way rely on these dynamics-dependent approaches to compute entropy. We only list them here to provide a broader context.

\section{Removing contractions with an arbitrary timelike vector}
\label{n-dependence}
	
Consider a regular point $P$ in  a spacetime with dimension $D\ge2$. We prove that if $f_{\mu\nu}$ is a symmetric tensor and for every timelike, unit, future-pointing vector $n^{\mu}$ it holds $f_{\mu\nu}n^{\mu}n^{\nu}=0$ in $P$, then $f_{\mu\nu}=0$. To carry out the proof, we introduce a local orthonormal coordinate system defined so that the metric locally reduces to the Minkowski one, i.e., $g_{\mu\nu}=\eta_{\mu\nu}$. We choose the local direction of time so that $n^{\mu}=\partial_{t}^{\mu}$ and denote the spatial coordinate vectors by $e_i^{\mu}=\partial_{x^i}^{\mu}$. Since $f_{\mu\nu}$ is a tensor, we can choose any coordinate system without loss of generality. Next, we define the following subset of unit timelike vectors in $P$
\begin{equation}
t^{\mu}_{ij}=\sqrt{\left(1+p^2+q^2\right)}n^{\mu}+pe_{i}^{\mu}+qe_{j}^{\mu},
\end{equation}
where $i$, $j$, are natural numbers such that $0<i<j\le n-1$, and $p$, $q$ are arbitrary real numbers. Since we require that $f_{\mu\nu}t^{\mu}_{ij}t^{\nu}_{ij}=0$ for every $t^{\mu}_{ij}$, then we have for any $p$, $q$ and any $i<j$
\begin{align}
\nonumber &\left(1+p^2+q^2\right)f_{00}+p^2f_{ii}+q^2f_{jj}+2p\sqrt{\left(1+p^2+q^2\right)}f_{0i} \\
&+2q\sqrt{\left(1+p^2+q^2\right)}f_{0j}+2pqf_{ij}=0. \label{f condition}
\end{align}
Thence, every coefficient in the expansion of the left hand side in the powers of $p$, $q$ must be zero. The first few conditions implied by this procedure are
\begin{align}
\nonumber f_{00}=&2pf_{0i}=2qf_{0j}=p^2\left(f_{00}+f_{ii}\right) \\
=&q^2\left(f_{00}+f_{jj}\right)=2pqf_{ij}=0.
\end{align}
To satisfy these requirements for every $i$, $j$, we must have $f_{\mu\nu}\left(P\right)=0$.

\bibliography{bibliography}

\begin{thebibliography}{98}%
\makeatletter
\providecommand \@ifxundefined [1]{%
 \@ifx{#1\undefined}
}%
\providecommand \@ifnum [1]{%
 \ifnum #1\expandafter \@firstoftwo
 \else \expandafter \@secondoftwo
 \fi
}%
\providecommand \@ifx [1]{%
 \ifx #1\expandafter \@firstoftwo
 \else \expandafter \@secondoftwo
 \fi
}%
\providecommand \natexlab [1]{#1}%
\providecommand \enquote  [1]{``#1''}%
\providecommand \bibnamefont  [1]{#1}%
\providecommand \bibfnamefont [1]{#1}%
\providecommand \citenamefont [1]{#1}%
\providecommand \href@noop [0]{\@secondoftwo}%
\providecommand \href [0]{\begingroup \@sanitize@url \@href}%
\providecommand \@href[1]{\@@startlink{#1}\@@href}%
\providecommand \@@href[1]{\endgroup#1\@@endlink}%
\providecommand \@sanitize@url [0]{\catcode `\\12\catcode `\$12\catcode `\&12\catcode `\#12\catcode `\^12\catcode `\_12\catcode `\%12\relax}%
\providecommand \@@startlink[1]{}%
\providecommand \@@endlink[0]{}%
\providecommand \url  [0]{\begingroup\@sanitize@url \@url }%
\providecommand \@url [1]{\endgroup\@href {#1}{\urlprefix }}%
\providecommand \urlprefix  [0]{URL }%
\providecommand \Eprint [0]{\href }%
\providecommand \doibase [0]{https://doi.org/}%
\providecommand \selectlanguage [0]{\@gobble}%
\providecommand \bibinfo  [0]{\@secondoftwo}%
\providecommand \bibfield  [0]{\@secondoftwo}%
\providecommand \translation [1]{[#1]}%
\providecommand \BibitemOpen [0]{}%
\providecommand \bibitemStop [0]{}%
\providecommand \bibitemNoStop [0]{.\EOS\space}%
\providecommand \EOS [0]{\spacefactor3000\relax}%
\providecommand \BibitemShut  [1]{\csname bibitem#1\endcsname}%
\let\auto@bib@innerbib\@empty
%</preamble>
\bibitem [{\citenamefont {Bekenstein}(1973)}]{Bekenstein:1973}%
  \BibitemOpen
  \bibfield  {author} {\bibinfo {author} {\bibfnamefont {J.~D.}\ \bibnamefont {Bekenstein}},\ }\bibfield  {title} {\bibinfo {title} {Black {H}oles and {E}ntropy},\ }\href {https://doi.org/10.1103/PhysRevD.7.2333} {\bibfield  {journal} {\bibinfo  {journal} {Phys. Rev. D}\ }\textbf {\bibinfo {volume} {7}},\ \bibinfo {pages} {2333} (\bibinfo {year} {1973})}\BibitemShut {NoStop}%
\bibitem [{\citenamefont {Hawking}(1975)}]{Hawking:1975}%
  \BibitemOpen
  \bibfield  {author} {\bibinfo {author} {\bibfnamefont {S.~W.}\ \bibnamefont {Hawking}},\ }\bibfield  {title} {\bibinfo {title} {Particle creation by black holes},\ }\href {https://doi.org/10.1007/BF02345020} {\bibfield  {journal} {\bibinfo  {journal} {Commun. Math. Phys.}\ }\textbf {\bibinfo {volume} {43}},\ \bibinfo {pages} {199} (\bibinfo {year} {1975})}\BibitemShut {NoStop}%
\bibitem [{\citenamefont {Braden}\ \emph {et~al.}(1990)\citenamefont {Braden}, \citenamefont {Brown}, \citenamefont {Whiting},\ and\ \citenamefont {York~Jr.}}]{Braden:1990}%
  \BibitemOpen
  \bibfield  {author} {\bibinfo {author} {\bibfnamefont {H.~W.}\ \bibnamefont {Braden}}, \bibinfo {author} {\bibfnamefont {J.~D.}\ \bibnamefont {Brown}}, \bibinfo {author} {\bibfnamefont {B.~F.}\ \bibnamefont {Whiting}},\ and\ \bibinfo {author} {\bibfnamefont {J.~W.}\ \bibnamefont {York~Jr.}},\ }\bibfield  {title} {\bibinfo {title} {{Charged black hole in a grand canonical ensemble}},\ }\href {https://doi.org/10.1103/PhysRevD.42.3376} {\bibfield  {journal} {\bibinfo  {journal} {Phys. Rev. D}\ }\textbf {\bibinfo {volume} {42}},\ \bibinfo {pages} {3376} (\bibinfo {year} {1990})}\BibitemShut {NoStop}%
\bibitem [{\citenamefont {Jacobson}(1995)}]{Jacobson:1995ab}%
  \BibitemOpen
  \bibfield  {author} {\bibinfo {author} {\bibfnamefont {T.}~\bibnamefont {Jacobson}},\ }\bibfield  {title} {\bibinfo {title} {Thermodynamics of space-time: The {E}instein equation of state},\ }\href {https://doi.org/10.1103/PhysRevLett.75.1260} {\bibfield  {journal} {\bibinfo  {journal} {Phys. Rev. Lett.}\ }\textbf {\bibinfo {volume} {75}},\ \bibinfo {pages} {1260} (\bibinfo {year} {1995})}\BibitemShut {NoStop}%
\bibitem [{\citenamefont {Jacobson}\ and\ \citenamefont {Parentani}(2003)}]{Jacobson:2003}%
  \BibitemOpen
  \bibfield  {author} {\bibinfo {author} {\bibfnamefont {T.}~\bibnamefont {Jacobson}}\ and\ \bibinfo {author} {\bibfnamefont {R.}~\bibnamefont {Parentani}},\ }\bibfield  {title} {\bibinfo {title} {Horizon entropy},\ }\href {https://doi.org/10.1023/A:1023785123428} {\bibfield  {journal} {\bibinfo  {journal} {Found. Phys.}\ }\textbf {\bibinfo {volume} {33}},\ \bibinfo {pages} {323} (\bibinfo {year} {2003})}\BibitemShut {NoStop}%
\bibitem [{\citenamefont {Padmanabhan}(2010)}]{Padmanabhan:2010}%
  \BibitemOpen
  \bibfield  {author} {\bibinfo {author} {\bibfnamefont {T.}~\bibnamefont {Padmanabhan}},\ }\bibfield  {title} {\bibinfo {title} {Thermodynamical aspects of gravity: New insights},\ }\href {https://doi.org/10.1088/0034-4885/73/4/046901} {\bibfield  {journal} {\bibinfo  {journal} {Rep. Prog. Phys.}\ }\textbf {\bibinfo {volume} {73}},\ \bibinfo {pages} {046901} (\bibinfo {year} {2010})}\BibitemShut {NoStop}%
\bibitem [{\citenamefont {Jacobson}(2016)}]{Jacobson:2015}%
  \BibitemOpen
  \bibfield  {author} {\bibinfo {author} {\bibfnamefont {T.}~\bibnamefont {Jacobson}},\ }\bibfield  {title} {\bibinfo {title} {Entanglement {E}quilibrium and the {E}instein equation},\ }\href {https://doi.org/10.1103/PhysRevLett.116.201101} {\bibfield  {journal} {\bibinfo  {journal} {Phys. Rev. Lett.}\ }\textbf {\bibinfo {volume} {116}},\ \bibinfo {pages} {201101} (\bibinfo {year} {2016})}\BibitemShut {NoStop}%
\bibitem [{\citenamefont {Bardeen}\ \emph {et~al.}(1973)\citenamefont {Bardeen}, \citenamefont {Carter},\ and\ \citenamefont {Hawking}}]{Bardeen:1973}%
  \BibitemOpen
  \bibfield  {author} {\bibinfo {author} {\bibfnamefont {J.~D.}\ \bibnamefont {Bardeen}}, \bibinfo {author} {\bibfnamefont {B.}~\bibnamefont {Carter}},\ and\ \bibinfo {author} {\bibfnamefont {S.~W.}\ \bibnamefont {Hawking}},\ }\bibfield  {title} {\bibinfo {title} {Black holes and entropy},\ }\href {https://doi.org/10.1007/BF01645742} {\bibfield  {journal} {\bibinfo  {journal} {Phys. Rev. D}\ }\textbf {\bibinfo {volume} {7}},\ \bibinfo {pages} {2333} (\bibinfo {year} {1973})}\BibitemShut {NoStop}%
\bibitem [{\citenamefont {Gibbons}\ and\ \citenamefont {Hawking}(1977)}]{Gibbons:1977}%
  \BibitemOpen
  \bibfield  {author} {\bibinfo {author} {\bibfnamefont {G.~W.}\ \bibnamefont {Gibbons}}\ and\ \bibinfo {author} {\bibfnamefont {S.~W.}\ \bibnamefont {Hawking}},\ }\bibfield  {title} {\bibinfo {title} {Action integrals and partition functions in quantum gravity},\ }\href {https://doi.org/10.1103/PhysRevD.15.2752} {\bibfield  {journal} {\bibinfo  {journal} {Phys. Rev. D}\ }\textbf {\bibinfo {volume} {15}},\ \bibinfo {pages} {2752} (\bibinfo {year} {1977})}\BibitemShut {NoStop}%
\bibitem [{\citenamefont {Wald}(1993)}]{Wald:1993}%
  \BibitemOpen
  \bibfield  {author} {\bibinfo {author} {\bibfnamefont {R.~M.}\ \bibnamefont {Wald}},\ }\bibfield  {title} {\bibinfo {title} {Black hole entropy is {N}oether charge},\ }\href {https://doi.org/10.1103/PhysRevD.48.R3427} {\bibfield  {journal} {\bibinfo  {journal} {Phys. Rev. D}\ }\textbf {\bibinfo {volume} {48}},\ \bibinfo {pages} {3427} (\bibinfo {year} {1993})}\BibitemShut {NoStop}%
\bibitem [{\citenamefont {Iyer}\ and\ \citenamefont {Wald}(1994)}]{Wald:1994}%
  \BibitemOpen
  \bibfield  {author} {\bibinfo {author} {\bibfnamefont {V.}~\bibnamefont {Iyer}}\ and\ \bibinfo {author} {\bibfnamefont {R.~M.}\ \bibnamefont {Wald}},\ }\bibfield  {title} {\bibinfo {title} {Some properties of {N}oether charge and a proposal for dynamical black hole entropy},\ }\href {https://doi.org/10.1103/PhysRevD.50.846} {\bibfield  {journal} {\bibinfo  {journal} {Phys. Rev. D}\ }\textbf {\bibinfo {volume} {50}},\ \bibinfo {pages} {846} (\bibinfo {year} {1994})}\BibitemShut {NoStop}%
\bibitem [{\citenamefont {Jacobson}\ and\ \citenamefont {Visser}(2019{\natexlab{a}})}]{Jacobson:2019}%
  \BibitemOpen
  \bibfield  {author} {\bibinfo {author} {\bibfnamefont {T.}~\bibnamefont {Jacobson}}\ and\ \bibinfo {author} {\bibfnamefont {M.~R.}\ \bibnamefont {Visser}},\ }\bibfield  {title} {\bibinfo {title} {Gravitational thermodynamics of causal diamonds in ({A})d{S}},\ }\href {https://doi.org/10.21468/SciPostPhys.7.6.079} {\bibfield  {journal} {\bibinfo  {journal} {SciPost Phys.}\ }\textbf {\bibinfo {volume} {7}},\ \bibinfo {pages} {079} (\bibinfo {year} {2019}{\natexlab{a}})}\BibitemShut {NoStop}%
\bibitem [{\citenamefont {Comp\`{e}re}\ and\ \citenamefont {Fiorucci}(2018)}]{Compere:2018}%
  \BibitemOpen
  \bibfield  {author} {\bibinfo {author} {\bibfnamefont {G.}~\bibnamefont {Comp\`{e}re}}\ and\ \bibinfo {author} {\bibfnamefont {A.}~\bibnamefont {Fiorucci}},\ }\href {https://doi.org/10.1007/978-3-030-04260-8} {\emph {\bibinfo {title} {Advanced Lectures on General Relativity}}}\ (\bibinfo  {publisher} {Springer International Publishing},\ \bibinfo {year} {2018})\BibitemShut {NoStop}%
\bibitem [{\citenamefont {Margalef-Bentabol}\ and\ \citenamefont {Villase\~nor}(2021)}]{Margalef:2021}%
  \BibitemOpen
  \bibfield  {author} {\bibinfo {author} {\bibfnamefont {J.}~\bibnamefont {Margalef-Bentabol}}\ and\ \bibinfo {author} {\bibfnamefont {E.~J.~S.}\ \bibnamefont {Villase\~nor}},\ }\bibfield  {title} {\bibinfo {title} {Geometric formulation of the covariant phase space methods with boundaries},\ }\href {https://doi.org/10.1103/PhysRevD.103.025011} {\bibfield  {journal} {\bibinfo  {journal} {Phys. Rev. D}\ }\textbf {\bibinfo {volume} {103}},\ \bibinfo {pages} {025011} (\bibinfo {year} {2021})}\BibitemShut {NoStop}%
\bibitem [{\citenamefont {Alonso-Serrano}\ \emph {et~al.}(2023)\citenamefont {Alonso-Serrano}, \citenamefont {Garay},\ and\ \citenamefont {Li\v{s}ka}}]{Alonso:2022}%
  \BibitemOpen
  \bibfield  {author} {\bibinfo {author} {\bibfnamefont {A.}~\bibnamefont {Alonso-Serrano}}, \bibinfo {author} {\bibfnamefont {L.~J.}\ \bibnamefont {Garay}},\ and\ \bibinfo {author} {\bibfnamefont {M.}~\bibnamefont {Li\v{s}ka}},\ }\bibfield  {title} {\bibinfo {title} {Noether charge formalism for {W}eyl transverse gravity},\ }\href {https://doi.org/10.1088/1361-6382/acace3} {\bibfield  {journal} {\bibinfo  {journal} {Class. Quant. Grav.}\ }\textbf {\bibinfo {volume} {40}},\ \bibinfo {pages} {025012} (\bibinfo {year} {2023})}\BibitemShut {NoStop}%
\bibitem [{\citenamefont {Alonso-Serrano}\ \emph {et~al.}(2022)\citenamefont {Alonso-Serrano}, \citenamefont {Garay},\ and\ \citenamefont {Li\v{s}ka}}]{Alonso:2022b}%
  \BibitemOpen
  \bibfield  {author} {\bibinfo {author} {\bibfnamefont {A.}~\bibnamefont {Alonso-Serrano}}, \bibinfo {author} {\bibfnamefont {L.~J.}\ \bibnamefont {Garay}},\ and\ \bibinfo {author} {\bibfnamefont {M.}~\bibnamefont {Li\v{s}ka}},\ }\bibfield  {title} {\bibinfo {title} {Noether charge formalism for {W}eyl invariant theories of gravity},\ }\href {https://doi.org/10.1103/PhysRevD.106.064024} {\bibfield  {journal} {\bibinfo  {journal} {Phys. Rev. D}\ }\textbf {\bibinfo {volume} {106}},\ \bibinfo {pages} {064024} (\bibinfo {year} {2022})}\BibitemShut {NoStop}%
\bibitem [{\citenamefont {Hollands}\ \emph {et~al.}(2024)\citenamefont {Hollands}, \citenamefont {Wald},\ and\ \citenamefont {Zhang}}]{Hollands:2024}%
  \BibitemOpen
  \bibfield  {author} {\bibinfo {author} {\bibfnamefont {S.}~\bibnamefont {Hollands}}, \bibinfo {author} {\bibfnamefont {R.~M.}\ \bibnamefont {Wald}},\ and\ \bibinfo {author} {\bibfnamefont {V.~G.}\ \bibnamefont {Zhang}},\ }\bibfield  {title} {\bibinfo {title} {Entropy of dynamical black holes},\ }\href {https://doi.org/10.1103/PhysRevD.110.024070} {\bibfield  {journal} {\bibinfo  {journal} {Phys. Rev. D}\ }\textbf {\bibinfo {volume} {110}},\ \bibinfo {pages} {024070} (\bibinfo {year} {2024})}\BibitemShut {NoStop}%
\bibitem [{\citenamefont {Padmanabhan}(2008)}]{Padmanabhan:2008}%
  \BibitemOpen
  \bibfield  {author} {\bibinfo {author} {\bibfnamefont {T.}~\bibnamefont {Padmanabhan}},\ }\bibfield  {title} {\bibinfo {title} {Dark energy and gravity},\ }\href {https://doi.org/10.1007/s10714-007-0555-7} {\bibfield  {journal} {\bibinfo  {journal} {Gen. Relativ. Gravit.}\ }\textbf {\bibinfo {volume} {40}},\ \bibinfo {pages} {529} (\bibinfo {year} {2008})}\BibitemShut {NoStop}%
\bibitem [{\citenamefont {Guedens}\ \emph {et~al.}(2012)\citenamefont {Guedens}, \citenamefont {Jacobson},\ and\ \citenamefont {Sarkar}}]{Jacobson:2012}%
  \BibitemOpen
  \bibfield  {author} {\bibinfo {author} {\bibfnamefont {R.}~\bibnamefont {Guedens}}, \bibinfo {author} {\bibfnamefont {T.}~\bibnamefont {Jacobson}},\ and\ \bibinfo {author} {\bibfnamefont {S.}~\bibnamefont {Sarkar}},\ }\bibfield  {title} {\bibinfo {title} {Horizon entropy and higher curvature equations of state},\ }\href {https://doi.org/10.1103/PhysRevD.85.064017} {\bibfield  {journal} {\bibinfo  {journal} {Phys. Rev. D}\ }\textbf {\bibinfo {volume} {85}},\ \bibinfo {pages} {064017} (\bibinfo {year} {2012})}\BibitemShut {NoStop}%
\bibitem [{\citenamefont {Bueno}\ \emph {et~al.}(2017)\citenamefont {Bueno}, \citenamefont {Min}, \citenamefont {Speranza},\ and\ \citenamefont {Visser}}]{Bueno:2017}%
  \BibitemOpen
  \bibfield  {author} {\bibinfo {author} {\bibfnamefont {P.}~\bibnamefont {Bueno}}, \bibinfo {author} {\bibfnamefont {V.~S.}\ \bibnamefont {Min}}, \bibinfo {author} {\bibfnamefont {A.~J.}\ \bibnamefont {Speranza}},\ and\ \bibinfo {author} {\bibfnamefont {M.~R.}\ \bibnamefont {Visser}},\ }\bibfield  {title} {\bibinfo {title} {Entanglement equilibrium for higher order gravity},\ }\href {https://doi.org/10.1103/PhysRevD.95.046003} {\bibfield  {journal} {\bibinfo  {journal} {Phys. Rev. D}\ }\textbf {\bibinfo {volume} {95}},\ \bibinfo {pages} {046003} (\bibinfo {year} {2017})}\BibitemShut {NoStop}%
\bibitem [{\citenamefont {Parikh}\ and\ \citenamefont {Svesko}(2018)}]{Svesko:2018}%
  \BibitemOpen
  \bibfield  {author} {\bibinfo {author} {\bibfnamefont {M.}~\bibnamefont {Parikh}}\ and\ \bibinfo {author} {\bibfnamefont {A.}~\bibnamefont {Svesko}},\ }\bibfield  {title} {\bibinfo {title} {Einstein's equations from the stretched future light cone},\ }\href {https://doi.org/10.1103/PhysRevD.98.026018} {\bibfield  {journal} {\bibinfo  {journal} {Phys. Rev. D}\ }\textbf {\bibinfo {volume} {98}},\ \bibinfo {pages} {026018} (\bibinfo {year} {2018})}\BibitemShut {NoStop}%
\bibitem [{\citenamefont {Svesko}(2019)}]{Svesko:2019}%
  \BibitemOpen
  \bibfield  {author} {\bibinfo {author} {\bibfnamefont {A.}~\bibnamefont {Svesko}},\ }\bibfield  {title} {\bibinfo {title} {Equilibrium to {E}instein: Entanglement, thermodynamics, and gravity},\ }\href {https://doi.org/10.1103/PhysRevD.99.086006} {\bibfield  {journal} {\bibinfo  {journal} {Phys. Rev. D}\ }\textbf {\bibinfo {volume} {99}},\ \bibinfo {pages} {086006} (\bibinfo {year} {2019})}\BibitemShut {NoStop}%
\bibitem [{\citenamefont {Bombelli}\ \emph {et~al.}(1986)\citenamefont {Bombelli}, \citenamefont {Koul}, \citenamefont {Lee},\ and\ \citenamefont {Sorkin}}]{Sorkin:1986}%
  \BibitemOpen
  \bibfield  {author} {\bibinfo {author} {\bibfnamefont {L.}~\bibnamefont {Bombelli}}, \bibinfo {author} {\bibfnamefont {R.~K.}\ \bibnamefont {Koul}}, \bibinfo {author} {\bibfnamefont {J.}~\bibnamefont {Lee}},\ and\ \bibinfo {author} {\bibfnamefont {R.~D.}\ \bibnamefont {Sorkin}},\ }\bibfield  {title} {\bibinfo {title} {Quantum source of entropy for black holes},\ }\href {https://doi.org/10.1103/PhysRevD.34.373} {\bibfield  {journal} {\bibinfo  {journal} {Phys. Rev. D}\ }\textbf {\bibinfo {volume} {34}},\ \bibinfo {pages} {373} (\bibinfo {year} {1986})}\BibitemShut {NoStop}%
\bibitem [{\citenamefont {Srednicki}(1993)}]{Srednicki:1993}%
  \BibitemOpen
  \bibfield  {author} {\bibinfo {author} {\bibfnamefont {M.}~\bibnamefont {Srednicki}},\ }\bibfield  {title} {\bibinfo {title} {Entropy and area},\ }\href {https://doi.org/10.1103/PhysRevLett.71.666} {\bibfield  {journal} {\bibinfo  {journal} {Phys. Rev. Lett.}\ }\textbf {\bibinfo {volume} {71}},\ \bibinfo {pages} {666} (\bibinfo {year} {1993})}\BibitemShut {NoStop}%
\bibitem [{\citenamefont {Solodukhin}(2011)}]{Solodukhin:2011}%
  \BibitemOpen
  \bibfield  {author} {\bibinfo {author} {\bibfnamefont {S.}~\bibnamefont {Solodukhin}},\ }\bibfield  {title} {\bibinfo {title} {Entanglement entropy of black holes},\ }\href {https://doi.org/10.12942/lrr-2011-8} {\bibfield  {journal} {\bibinfo  {journal} {Living Rev. Rel.}\ }\textbf {\bibinfo {volume} {214}},\ \bibinfo {pages} {8} (\bibinfo {year} {2011})}\BibitemShut {NoStop}%
\bibitem [{\citenamefont {Chirco}\ and\ \citenamefont {Liberati}(2010)}]{Chirco:2010}%
  \BibitemOpen
  \bibfield  {author} {\bibinfo {author} {\bibfnamefont {G.}~\bibnamefont {Chirco}}\ and\ \bibinfo {author} {\bibfnamefont {S.}~\bibnamefont {Liberati}},\ }\bibfield  {title} {\bibinfo {title} {Non-equilibrium thermodynamics of spacetime: the role of gravitational dissipation},\ }\href {https://doi.org/10.1103/PhysRevD.81.024016} {\bibfield  {journal} {\bibinfo  {journal} {Phys. Rev. D}\ }\textbf {\bibinfo {volume} {81}},\ \bibinfo {pages} {024016} (\bibinfo {year} {2010})}\BibitemShut {NoStop}%
\bibitem [{\citenamefont {Eling}\ \emph {et~al.}(2006)\citenamefont {Eling}, \citenamefont {Guedens},\ and\ \citenamefont {Jacobson}}]{Eling:2006}%
  \BibitemOpen
  \bibfield  {author} {\bibinfo {author} {\bibfnamefont {C.}~\bibnamefont {Eling}}, \bibinfo {author} {\bibfnamefont {R.}~\bibnamefont {Guedens}},\ and\ \bibinfo {author} {\bibfnamefont {T.}~\bibnamefont {Jacobson}},\ }\bibfield  {title} {\bibinfo {title} {Non-equilibrium thermodynamics of spacetime},\ }\href {https://doi.org/PhysRevLett.96.121301} {\bibfield  {journal} {\bibinfo  {journal} {Phys. Rev. Lett.}\ }\textbf {\bibinfo {volume} {96}},\ \bibinfo {pages} {121301} (\bibinfo {year} {2006})}\BibitemShut {NoStop}%
\bibitem [{\citenamefont {Di~Casola}\ \emph {et~al.}(2014)\citenamefont {Di~Casola}, \citenamefont {Liberati},\ and\ \citenamefont {Sonego}}]{Casola:2014}%
  \BibitemOpen
  \bibfield  {author} {\bibinfo {author} {\bibfnamefont {E.}~\bibnamefont {Di~Casola}}, \bibinfo {author} {\bibfnamefont {S.}~\bibnamefont {Liberati}},\ and\ \bibinfo {author} {\bibfnamefont {S.}~\bibnamefont {Sonego}},\ }\bibfield  {title} {\bibinfo {title} {Weak equivalence principle for self-gravitating bodies: A sieve for purely metric theories of gravity},\ }\href {https://doi.org/10.1103/PhysRevD.89.084053} {\bibfield  {journal} {\bibinfo  {journal} {Phys. Rev. D}\ }\textbf {\bibinfo {volume} {89}},\ \bibinfo {pages} {084053} (\bibinfo {year} {2014})}\BibitemShut {NoStop}%
\bibitem [{\citenamefont {\'{A}lvarez}\ \emph {et~al.}(2006)\citenamefont {\'{A}lvarez}, \citenamefont {Blas}, \citenamefont {Garriga},\ and\ \citenamefont {Verdaguer}}]{Alvarez:2006}%
  \BibitemOpen
  \bibfield  {author} {\bibinfo {author} {\bibfnamefont {E.}~\bibnamefont {\'{A}lvarez}}, \bibinfo {author} {\bibfnamefont {D.}~\bibnamefont {Blas}}, \bibinfo {author} {\bibfnamefont {J.}~\bibnamefont {Garriga}},\ and\ \bibinfo {author} {\bibfnamefont {E.}~\bibnamefont {Verdaguer}},\ }\bibfield  {title} {\bibinfo {title} {{Transverse {F}ierz–{P}auli symmetry}},\ }\href {https://doi.org/10.1016/j.nuclphysb.2006.08.003} {\bibfield  {journal} {\bibinfo  {journal} {Nucl. Phys. B.}\ }\textbf {\bibinfo {volume} {756}},\ \bibinfo {pages} {148} (\bibinfo {year} {2006})}\BibitemShut {NoStop}%
\bibitem [{\citenamefont {Carballo-Rubio}\ \emph {et~al.}(2022)\citenamefont {Carballo-Rubio}, \citenamefont {Garay},\ and\ \citenamefont {Garc\'{i}a-Moreno}}]{Carballo:2022}%
  \BibitemOpen
  \bibfield  {author} {\bibinfo {author} {\bibfnamefont {R.}~\bibnamefont {Carballo-Rubio}}, \bibinfo {author} {\bibfnamefont {L.~J.}\ \bibnamefont {Garay}},\ and\ \bibinfo {author} {\bibfnamefont {G.}~\bibnamefont {Garc\'{i}a-Moreno}},\ }\bibfield  {title} {\bibinfo {title} {Unimodular gravity vs general relativity: A status report},\ }\href {https://doi.org/10.1088/1361-6382/aca386} {\bibfield  {journal} {\bibinfo  {journal} {Class. Quant. Grav.}\ }\textbf {\bibinfo {volume} {39}},\ \bibinfo {pages} {243001} (\bibinfo {year} {2022})}\BibitemShut {NoStop}%
\bibitem [{\citenamefont {\'{A}lvarez}\ \emph {et~al.}(2023)\citenamefont {\'{A}lvarez}, \citenamefont {Anero},\ and\ \citenamefont {Sanchez-Ruiz}}]{Alvarez:2023}%
  \BibitemOpen
  \bibfield  {author} {\bibinfo {author} {\bibfnamefont {E.}~\bibnamefont {\'{A}lvarez}}, \bibinfo {author} {\bibfnamefont {J.}~\bibnamefont {Anero}},\ and\ \bibinfo {author} {\bibfnamefont {I.}~\bibnamefont {Sanchez-Ruiz}},\ }\bibfield  {title} {\bibinfo {title} {{Physical charges versus conformal invariance in unimodular gravity}},\ }\href {https://doi.org/10.1142/S0217751X23501324} {\bibfield  {journal} {\bibinfo  {journal} {Int. J. Mod. Phys. A}\ }\textbf {\bibinfo {volume} {38}},\ \bibinfo {pages} {2350132} (\bibinfo {year} {2023})}\BibitemShut {NoStop}%
\bibitem [{\citenamefont {Henneaux}\ and\ \citenamefont {Teitelboim}(1989)}]{Henneaux:1989}%
  \BibitemOpen
  \bibfield  {author} {\bibinfo {author} {\bibfnamefont {M.}~\bibnamefont {Henneaux}}\ and\ \bibinfo {author} {\bibfnamefont {C.}~\bibnamefont {Teitelboim}},\ }\bibfield  {title} {\bibinfo {title} {The {C}osmological {C}onstant and {G}eneral {C}ovariance},\ }\href {https://doi.org/10.1016/0370-2693(89)91251-3} {\bibfield  {journal} {\bibinfo  {journal} {Phys. Lett. B}\ }\textbf {\bibinfo {volume} {222}},\ \bibinfo {pages} {195} (\bibinfo {year} {1989})}\BibitemShut {NoStop}%
\bibitem [{\citenamefont {Padilla}\ and\ \citenamefont {Saltas}(2014)}]{Padilla:2014}%
  \BibitemOpen
  \bibfield  {author} {\bibinfo {author} {\bibfnamefont {A.}~\bibnamefont {Padilla}}\ and\ \bibinfo {author} {\bibfnamefont {I.~D.}\ \bibnamefont {Saltas}},\ }\bibfield  {title} {\bibinfo {title} {A note on classical and quantum unimodular gravity},\ }\href {https://doi.org/10.1140/epjc/s10052-015-3767-0} {\bibfield  {journal} {\bibinfo  {journal} {Eur. Phys. J. C}\ }\textbf {\bibinfo {volume} {75}},\ \bibinfo {pages} {561} (\bibinfo {year} {2014})}\BibitemShut {NoStop}%
\bibitem [{\citenamefont {Bufalo}\ \emph {et~al.}(2015)\citenamefont {Bufalo}, \citenamefont {Oksanen},\ and\ \citenamefont {Tureanu}}]{Bufalo:2015}%
  \BibitemOpen
  \bibfield  {author} {\bibinfo {author} {\bibfnamefont {R.}~\bibnamefont {Bufalo}}, \bibinfo {author} {\bibfnamefont {M.}~\bibnamefont {Oksanen}},\ and\ \bibinfo {author} {\bibfnamefont {A.}~\bibnamefont {Tureanu}},\ }\bibfield  {title} {\bibinfo {title} {How unimodular gravity theories differ from general relativity at quantum level},\ }\href {https://doi.org/10.1140/epjc/s10052-015-3683-3} {\bibfield  {journal} {\bibinfo  {journal} {Eur. Phys. J. C}\ }\textbf {\bibinfo {volume} {75}},\ \bibinfo {pages} {477} (\bibinfo {year} {2015})}\BibitemShut {NoStop}%
\bibitem [{\citenamefont {Barcel\'{o}}\ \emph {et~al.}(2014)\citenamefont {Barcel\'{o}}, \citenamefont {Carballo-Rubio},\ and\ \citenamefont {Garay}}]{Barcelo:2014}%
  \BibitemOpen
  \bibfield  {author} {\bibinfo {author} {\bibfnamefont {C.}~\bibnamefont {Barcel\'{o}}}, \bibinfo {author} {\bibfnamefont {R.}~\bibnamefont {Carballo-Rubio}},\ and\ \bibinfo {author} {\bibfnamefont {L.~J.}\ \bibnamefont {Garay}},\ }\bibfield  {title} {\bibinfo {title} {Unimodular gravity and general relativity from graviton self-interactions},\ }\href {https://doi.org/10.1103/PhysRevD.89.124019} {\bibfield  {journal} {\bibinfo  {journal} {Phys. Rev. D}\ }\textbf {\bibinfo {volume} {89}},\ \bibinfo {pages} {124019} (\bibinfo {year} {2014})}\BibitemShut {NoStop}%
\bibitem [{\citenamefont {\'{A}lvarez}\ \emph {et~al.}(2016)\citenamefont {\'{A}lvarez}, \citenamefont {Gonz\'{a}lez-Mart\'{\i}n},\ and\ \citenamefont {Mart\'{\i}n}}]{Alvarez:2016}%
  \BibitemOpen
  \bibfield  {author} {\bibinfo {author} {\bibfnamefont {E.}~\bibnamefont {\'{A}lvarez}}, \bibinfo {author} {\bibfnamefont {S.}~\bibnamefont {Gonz\'{a}lez-Mart\'{\i}n}},\ and\ \bibinfo {author} {\bibfnamefont {C.~P.}\ \bibnamefont {Mart\'{\i}n}},\ }\bibfield  {title} {\bibinfo {title} {Note on the gauge symmetries of unimodular gravity},\ }\href {https://doi.org/10.1103/PhysRevD.93.123018} {\bibfield  {journal} {\bibinfo  {journal} {Phys. Rev. D}\ }\textbf {\bibinfo {volume} {93}},\ \bibinfo {pages} {123018} (\bibinfo {year} {2016})}\BibitemShut {NoStop}%
\bibitem [{\citenamefont {Barcel\'{o}}\ \emph {et~al.}(2018)\citenamefont {Barcel\'{o}}, \citenamefont {Garay},\ and\ \citenamefont {Carballo-Rubio}}]{Barcelo:2018}%
  \BibitemOpen
  \bibfield  {author} {\bibinfo {author} {\bibfnamefont {C.}~\bibnamefont {Barcel\'{o}}}, \bibinfo {author} {\bibfnamefont {L.~J.}\ \bibnamefont {Garay}},\ and\ \bibinfo {author} {\bibfnamefont {R.}~\bibnamefont {Carballo-Rubio}},\ }\bibfield  {title} {\bibinfo {title} {Absence of cosmological constant problem in special relativistic field theory of gravity},\ }\href {https://doi.org/10.1016/j.aop.2018.08.016} {\bibfield  {journal} {\bibinfo  {journal} {Annals of Physics}\ }\textbf {\bibinfo {volume} {398}},\ \bibinfo {pages} {9} (\bibinfo {year} {2018})}\BibitemShut {NoStop}%
\bibitem [{\citenamefont {Tiwari}()}]{Tiwari:2006}%
  \BibitemOpen
  \bibfield  {author} {\bibinfo {author} {\bibfnamefont {S.~C.}\ \bibnamefont {Tiwari}},\ }\href {https://doi.org/10.48550/arXiv.gr-qc/0612099} {\bibinfo {title} {Thermodynamics of spacetime and unimodular relativity}},\ \Eprint {https://arxiv.org/abs/0612099} {arxiv:0612099 [gr-qc]} \BibitemShut {NoStop}%
\bibitem [{\citenamefont {Alonso-Serrano}\ and\ \citenamefont {Li\v{s}ka}(2020)}]{Alonso:2020}%
  \BibitemOpen
  \bibfield  {author} {\bibinfo {author} {\bibfnamefont {A.}~\bibnamefont {Alonso-Serrano}}\ and\ \bibinfo {author} {\bibfnamefont {M.}~\bibnamefont {Li\v{s}ka}},\ }\bibfield  {title} {\bibinfo {title} {New perspective on thermodynamics of spacetime: The emergence of unimodular gravity and the equivalence of entropies},\ }\href {https://doi.org/10.1103/PhysRevD.102.104056} {\bibfield  {journal} {\bibinfo  {journal} {Phys. Rev. D}\ }\textbf {\bibinfo {volume} {102}},\ \bibinfo {pages} {104056} (\bibinfo {year} {2020})}\BibitemShut {NoStop}%
\bibitem [{\citenamefont {Alonso-Serrano}\ and\ \citenamefont {Li\v{s}ka}(2022)}]{Alonso:2021}%
  \BibitemOpen
  \bibfield  {author} {\bibinfo {author} {\bibfnamefont {A.}~\bibnamefont {Alonso-Serrano}}\ and\ \bibinfo {author} {\bibfnamefont {M.}~\bibnamefont {Li\v{s}ka}},\ }\bibfield  {title} {\bibinfo {title} {Thermodynamics of spacetime and unimodular gravity},\ }\href {https://doi.org/10.1142/S0219887822300021} {\bibfield  {journal} {\bibinfo  {journal} {Int. J. Geom. Methods Mod. Phys.}\ }\textbf {\bibinfo {volume} {19}},\ \bibinfo {pages} {2230002} (\bibinfo {year} {2022})}\BibitemShut {NoStop}%
\bibitem [{\citenamefont {Alonso-Serrano}\ \emph {et~al.}()\citenamefont {Alonso-Serrano}, \citenamefont {Garay},\ and\ \citenamefont {Li\v{s}ka}}]{equivalence}%
  \BibitemOpen
  \bibfield  {author} {\bibinfo {author} {\bibfnamefont {A.}~\bibnamefont {Alonso-Serrano}}, \bibinfo {author} {\bibfnamefont {L.~J.}\ \bibnamefont {Garay}},\ and\ \bibinfo {author} {\bibfnamefont {M.}~\bibnamefont {Li\v{s}ka}},\ }\href {https://doi.org/1} {\bibinfo {title} {Equivalence principle(s) in {W}eyl transverse gravity}},\ \Eprint {https://arxiv.org/abs/1} {arxiv:1 [gr-qc]} \BibitemShut {NoStop}%
\bibitem [{\citenamefont {Verlinde}(2011)}]{Verlinde:2011}%
  \BibitemOpen
  \bibfield  {author} {\bibinfo {author} {\bibfnamefont {E.~P.}\ \bibnamefont {Verlinde}},\ }\bibfield  {title} {\bibinfo {title} {On the origin of gravity and the laws of {N}ewton},\ }\href {https://doi.org/10.1007/JHEP04(2011)029} {\bibfield  {journal} {\bibinfo  {journal} {J. High Energ. Phys.}\ }\textbf {\bibinfo {volume} {2011}}\bibinfo  {number} { (29)}}\BibitemShut {NoStop}%
\bibitem [{\citenamefont {Misner}\ \emph {et~al.}(2017)\citenamefont {Misner}, \citenamefont {Thorne},\ and\ \citenamefont {Wheeler}}]{MTW}%
  \BibitemOpen
\bibfield  {number} {  }\bibfield  {author} {\bibinfo {author} {\bibfnamefont {C.~W.}\ \bibnamefont {Misner}}, \bibinfo {author} {\bibfnamefont {K.~S.}\ \bibnamefont {Thorne}},\ and\ \bibinfo {author} {\bibfnamefont {J.~A.}\ \bibnamefont {Wheeler}},\ }\href@noop {} {\emph {\bibinfo {title} {Gravitation}}}\ (\bibinfo  {publisher} {Princeton University Press},\ \bibinfo {address} {Princeton},\ \bibinfo {year} {2017})\BibitemShut {NoStop}%
\bibitem [{\citenamefont {\'{A}lvarez}\ and\ \citenamefont {Herrero-Valea}(2013)}]{Alvarez:2013}%
  \BibitemOpen
  \bibfield  {author} {\bibinfo {author} {\bibfnamefont {E.}~\bibnamefont {\'{A}lvarez}}\ and\ \bibinfo {author} {\bibfnamefont {M.}~\bibnamefont {Herrero-Valea}},\ }\bibfield  {title} {\bibinfo {title} {{Unimodular gravity with external sources}},\ }\href {https://doi.org/10.1088/1475-7516/2013/01/014} {\bibfield  {journal} {\bibinfo  {journal} {J. Cosmol. Astropart. Phys.}\ }\textbf {\bibinfo {volume} {2013}}\bibinfo  {number} { (014)}}\BibitemShut {NoStop}%
\bibitem [{\citenamefont {Carballo-Rubio}(2015)}]{Carballo:2015}%
  \BibitemOpen
\bibfield  {number} {  }\bibfield  {author} {\bibinfo {author} {\bibfnamefont {R.}~\bibnamefont {Carballo-Rubio}},\ }\bibfield  {title} {\bibinfo {title} {Longitudinal diffeomorphisms obstruct the protection of vacuum energy},\ }\href {https://doi.org/10.1103/PhysRevD.91.124071} {\bibfield  {journal} {\bibinfo  {journal} {Phys. Rev. D}\ }\textbf {\bibinfo {volume} {91}},\ \bibinfo {pages} {124071} (\bibinfo {year} {2015})}\BibitemShut {NoStop}%
\bibitem [{\citenamefont {Di~Casola}\ \emph {et~al.}(2015)\citenamefont {Di~Casola}, \citenamefont {Liberati},\ and\ \citenamefont {Sonego}}]{Casola:2015}%
  \BibitemOpen
  \bibfield  {author} {\bibinfo {author} {\bibfnamefont {E.}~\bibnamefont {Di~Casola}}, \bibinfo {author} {\bibfnamefont {S.}~\bibnamefont {Liberati}},\ and\ \bibinfo {author} {\bibfnamefont {S.}~\bibnamefont {Sonego}},\ }\bibfield  {title} {\bibinfo {title} {Nonequivalence of equivalence principles},\ }\href {https://doi.org/10.1119/1.4895342} {\bibfield  {journal} {\bibinfo  {journal} {Am. J. Phys.}\ }\textbf {\bibinfo {volume} {20}},\ \bibinfo {pages} {39} (\bibinfo {year} {2015})}\BibitemShut {NoStop}%
\bibitem [{\citenamefont {Carroll}\ and\ \citenamefont {Remmen}(2016)}]{Carroll:2016}%
  \BibitemOpen
  \bibfield  {author} {\bibinfo {author} {\bibfnamefont {S.~M.}\ \bibnamefont {Carroll}}\ and\ \bibinfo {author} {\bibfnamefont {G.~N.}\ \bibnamefont {Remmen}},\ }\bibfield  {title} {\bibinfo {title} {What is the entropy in entropic gravity?},\ }\href {https://doi.org/10.1103/PhysRevD.93.124052} {\bibfield  {journal} {\bibinfo  {journal} {Phys. Rev. D}\ }\textbf {\bibinfo {volume} {93}},\ \bibinfo {pages} {124052} (\bibinfo {year} {2016})}\BibitemShut {NoStop}%
\bibitem [{\citenamefont {Mead}(1964)}]{Mead:1964}%
  \BibitemOpen
  \bibfield  {author} {\bibinfo {author} {\bibfnamefont {C.~A.}\ \bibnamefont {Mead}},\ }\bibfield  {title} {\bibinfo {title} {Possible connection between gravitation and fundamental length},\ }\href {https://doi.org/10.1103/PhysRev.135.B849} {\bibfield  {journal} {\bibinfo  {journal} {Phys. Rev.}\ }\textbf {\bibinfo {volume} {135}},\ \bibinfo {pages} {B849} (\bibinfo {year} {1964})}\BibitemShut {NoStop}%
\bibitem [{\citenamefont {Garay}(1995)}]{Garay:1994en}%
  \BibitemOpen
  \bibfield  {author} {\bibinfo {author} {\bibfnamefont {L.~J.}\ \bibnamefont {Garay}},\ }\bibfield  {title} {\bibinfo {title} {Quantum gravity and minimum length},\ }\href {https://doi.org/10.1142/S0217751X95000085} {\bibfield  {journal} {\bibinfo  {journal} {Int. J. Mod. Phys. A}\ }\textbf {\bibinfo {volume} {10}},\ \bibinfo {pages} {145} (\bibinfo {year} {1995})}\BibitemShut {NoStop}%
\bibitem [{\citenamefont {Hossenfelder}(2013)}]{Hossenfelder:2013}%
  \BibitemOpen
  \bibfield  {author} {\bibinfo {author} {\bibfnamefont {S.}~\bibnamefont {Hossenfelder}},\ }\bibfield  {title} {\bibinfo {title} {Minimal length scale scenarios for quantum gravity},\ }\href {https://doi.org/10.12942/lrr-2013-2} {\bibfield  {journal} {\bibinfo  {journal} {Living Rev. Relativ.}\ }\textbf {\bibinfo {volume} {16}},\ \bibinfo {pages} {2} (\bibinfo {year} {2013})}\BibitemShut {NoStop}%
\bibitem [{\citenamefont {Wang}(2019)}]{Wang:2019}%
  \BibitemOpen
  \bibfield  {author} {\bibinfo {author} {\bibfnamefont {J.}~\bibnamefont {Wang}},\ }\bibfield  {title} {\bibinfo {title} {Geometry of small causal diamonds},\ }\href {https://doi.org/10.1103/PhysRevD.100.064020} {\bibfield  {journal} {\bibinfo  {journal} {Phys. Rev. D}\ }\textbf {\bibinfo {volume} {100}},\ \bibinfo {pages} {064020} (\bibinfo {year} {2019})}\BibitemShut {NoStop}%
\bibitem [{\citenamefont {Chakraborty}\ \emph {et~al.}(2016)\citenamefont {Chakraborty}, \citenamefont {Bhattacharya},\ and\ \citenamefont {Padmanabhan}}]{Chakraborty:2016}%
  \BibitemOpen
  \bibfield  {author} {\bibinfo {author} {\bibfnamefont {S.}~\bibnamefont {Chakraborty}}, \bibinfo {author} {\bibfnamefont {S.}~\bibnamefont {Bhattacharya}},\ and\ \bibinfo {author} {\bibfnamefont {T.}~\bibnamefont {Padmanabhan}},\ }\bibfield  {title} {\bibinfo {title} {Entropy of a generic null surface from its associated {V}irasoro algebra},\ }\href {https://doi.org/10.1016/j.physletb.2016.10.059} {\bibfield  {journal} {\bibinfo  {journal} {Phys. Lett. B}\ }\textbf {\bibinfo {volume} {763}},\ \bibinfo {pages} {347} (\bibinfo {year} {2016})}\BibitemShut {NoStop}%
\bibitem [{\citenamefont {Jacobson}\ and\ \citenamefont {Visser}(2023{\natexlab{a}})}]{Jacobson:2023}%
  \BibitemOpen
  \bibfield  {author} {\bibinfo {author} {\bibfnamefont {T.}~\bibnamefont {Jacobson}}\ and\ \bibinfo {author} {\bibfnamefont {M.~R.}\ \bibnamefont {Visser}},\ }\bibfield  {title} {\bibinfo {title} {Partition function for a volume of space},\ }\href {https://doi.org/10.1103/PhysRevLett.130.221501} {\bibfield  {journal} {\bibinfo  {journal} {Phys. Rev. Lett.}\ }\textbf {\bibinfo {volume} {130}},\ \bibinfo {pages} {221501} (\bibinfo {year} {2023}{\natexlab{a}})}\BibitemShut {NoStop}%
\bibitem [{\citenamefont {Jacobson}\ and\ \citenamefont {Visser}(2023{\natexlab{b}})}]{Jacobson:2023b}%
  \BibitemOpen
  \bibfield  {author} {\bibinfo {author} {\bibfnamefont {T.}~\bibnamefont {Jacobson}}\ and\ \bibinfo {author} {\bibfnamefont {M.~R.}\ \bibnamefont {Visser}},\ }\bibfield  {title} {\bibinfo {title} {Entropy of causal diamond ensembles},\ }\href {https://doi.org/10.21468/SciPostPhys.15.1.023} {\bibfield  {journal} {\bibinfo  {journal} {SciPost Phys.}\ }\textbf {\bibinfo {volume} {15}},\ \bibinfo {pages} {023} (\bibinfo {year} {2023}{\natexlab{b}})}\BibitemShut {NoStop}%
\bibitem [{\citenamefont {Jensen}\ \emph {et~al.}(2023)\citenamefont {Jensen}, \citenamefont {Sorce},\ and\ \citenamefont {Speranza}}]{Speranza:2023}%
  \BibitemOpen
  \bibfield  {author} {\bibinfo {author} {\bibfnamefont {K.}~\bibnamefont {Jensen}}, \bibinfo {author} {\bibfnamefont {J.}~\bibnamefont {Sorce}},\ and\ \bibinfo {author} {\bibfnamefont {A.}~\bibnamefont {Speranza}},\ }\bibfield  {title} {\bibinfo {title} {Generalized entropy for general subregions in quantum gravity},\ }\href {https://doi.org/10.1007/JHEP12(2023)020} {\bibfield  {journal} {\bibinfo  {journal} {J. High Energ. Phys.}\ }\textbf {\bibinfo {volume} {2023}}\bibinfo  {number} { (20)}}\BibitemShut {NoStop}%
\bibitem [{\citenamefont {Belgiorno}\ and\ \citenamefont {Liberati}(1996)}]{Belgiorno:1996}%
  \BibitemOpen
\bibfield  {number} {  }\bibfield  {author} {\bibinfo {author} {\bibfnamefont {F.}~\bibnamefont {Belgiorno}}\ and\ \bibinfo {author} {\bibfnamefont {S.}~\bibnamefont {Liberati}},\ }\bibfield  {title} {\bibinfo {title} {Divergence problem in the black hole brick-wall model},\ }\href {https://doi.org/10.1103/PhysRevD.53.3172} {\bibfield  {journal} {\bibinfo  {journal} {Phys. Rev. D}\ }\textbf {\bibinfo {volume} {53}},\ \bibinfo {pages} {3172} (\bibinfo {year} {1996})}\BibitemShut {NoStop}%
\bibitem [{\citenamefont {Liberati}(1997)}]{Liberati:1997}%
  \BibitemOpen
  \bibfield  {author} {\bibinfo {author} {\bibfnamefont {S.}~\bibnamefont {Liberati}},\ }\bibfield  {title} {\bibinfo {title} {Problems in black-hole entropy interpretation},\ }\href {https://doi.org/10.48550/arXiv.gr-qc/9601032} {\bibfield  {journal} {\bibinfo  {journal} {Nuov. Cim. B}\ }\textbf {\bibinfo {volume} {112}},\ \bibinfo {pages} {405} (\bibinfo {year} {1997})}\BibitemShut {NoStop}%
\bibitem [{\citenamefont {Solodukhin}(2010)}]{Solodukhin:2010}%
  \BibitemOpen
  \bibfield  {author} {\bibinfo {author} {\bibfnamefont {S.~N.}\ \bibnamefont {Solodukhin}},\ }\bibfield  {title} {\bibinfo {title} {Entanglement entropy of round spheres},\ }\href {https://doi.org/10.1016/j.physletb.2010.09.018} {\bibfield  {journal} {\bibinfo  {journal} {Phys. Lett. B}\ }\textbf {\bibinfo {volume} {693}},\ \bibinfo {pages} {605} (\bibinfo {year} {2010})}\BibitemShut {NoStop}%
\bibitem [{\citenamefont {Baccetti}\ and\ \citenamefont {Visser}(2014)}]{Baccetti:2013ica}%
  \BibitemOpen
  \bibfield  {author} {\bibinfo {author} {\bibfnamefont {V.}~\bibnamefont {Baccetti}}\ and\ \bibinfo {author} {\bibfnamefont {M.}~\bibnamefont {Visser}},\ }\bibfield  {title} {\bibinfo {title} {Clausius entropy for arbitrary bifurcate null surfaces},\ }\href {https://doi.org/10.1088/0264-9381/31/3/035009} {\bibfield  {journal} {\bibinfo  {journal} {Class. Quant. Grav}\ }\textbf {\bibinfo {volume} {31}},\ \bibinfo {pages} {035009} (\bibinfo {year} {2014})}\BibitemShut {NoStop}%
\bibitem [{\citenamefont {Fulling}(1973)}]{Fulling:1973}%
  \BibitemOpen
  \bibfield  {author} {\bibinfo {author} {\bibfnamefont {S.~A.}\ \bibnamefont {Fulling}},\ }\bibfield  {title} {\bibinfo {title} {Nonuniqueness of canonical field quantization in {R}iemannian space-time},\ }\href {https://doi.org/10.1103/PhysRevD.7.2850} {\bibfield  {journal} {\bibinfo  {journal} {Phys. Rev. D}\ }\textbf {\bibinfo {volume} {7}},\ \bibinfo {pages} {2850} (\bibinfo {year} {1973})}\BibitemShut {NoStop}%
\bibitem [{\citenamefont {Bisognano}\ and\ \citenamefont {Wichmann}(1976)}]{Bisognano:1976}%
  \BibitemOpen
  \bibfield  {author} {\bibinfo {author} {\bibfnamefont {J.~J.}\ \bibnamefont {Bisognano}}\ and\ \bibinfo {author} {\bibfnamefont {E.~H.}\ \bibnamefont {Wichmann}},\ }\bibfield  {title} {\bibinfo {title} {On the duality condition for quantum fields},\ }\href {https://doi.org/10.1063/1.522898} {\bibfield  {journal} {\bibinfo  {journal} {J. Math. Phys.}\ }\textbf {\bibinfo {volume} {17}},\ \bibinfo {pages} {303} (\bibinfo {year} {1976})}\BibitemShut {NoStop}%
\bibitem [{\citenamefont {Unruh}(1976)}]{Unruh:1976}%
  \BibitemOpen
  \bibfield  {author} {\bibinfo {author} {\bibfnamefont {W.~G.}\ \bibnamefont {Unruh}},\ }\bibfield  {title} {\bibinfo {title} {Notes on black hole evaporation},\ }\href {https://doi.org/10.1103/PhysRevD.14.870} {\bibfield  {journal} {\bibinfo  {journal} {Phys.Rev. D}\ }\textbf {\bibinfo {volume} {14}},\ \bibinfo {pages} {870} (\bibinfo {year} {1976})}\BibitemShut {NoStop}%
\bibitem [{\citenamefont {Barbado}\ and\ \citenamefont {Visser}(2012)}]{Barbado:2012}%
  \BibitemOpen
  \bibfield  {author} {\bibinfo {author} {\bibfnamefont {L.~C.}\ \bibnamefont {Barbado}}\ and\ \bibinfo {author} {\bibfnamefont {M.}~\bibnamefont {Visser}},\ }\bibfield  {title} {\bibinfo {title} {Unruh-{D}e{W}itt detector event rate for trajectories with time-dependent acceleration},\ }\href {https://doi.org/10.1103/PhysRevD.86.084011} {\bibfield  {journal} {\bibinfo  {journal} {Phys. Rev. D}\ }\textbf {\bibinfo {volume} {86}},\ \bibinfo {pages} {084011} (\bibinfo {year} {2012})}\BibitemShut {NoStop}%
\bibitem [{\citenamefont {Perche}(2021)}]{Perche:2021}%
  \BibitemOpen
  \bibfield  {author} {\bibinfo {author} {\bibfnamefont {T.~R.}\ \bibnamefont {Perche}},\ }\bibfield  {title} {\bibinfo {title} {General features of the thermalization of particle detectors and the unruh effect},\ }\href {https://doi.org/10.1103/PhysRevD.104.065001} {\bibfield  {journal} {\bibinfo  {journal} {Phys. Rev. D}\ }\textbf {\bibinfo {volume} {104}},\ \bibinfo {pages} {065001} (\bibinfo {year} {2021})}\BibitemShut {NoStop}%
\bibitem [{\citenamefont {Perche}(2022)}]{Perche:2022b}%
  \BibitemOpen
  \bibfield  {author} {\bibinfo {author} {\bibfnamefont {T.~R.}\ \bibnamefont {Perche}},\ }\bibfield  {title} {\bibinfo {title} {Localized nonrelativistic quantum systems in curved spacetimes: A general characterization of particle detector models},\ }\href {https://doi.org/10.1103/PhysRevD.106.025018} {\bibfield  {journal} {\bibinfo  {journal} {Phys. Rev. D}\ }\textbf {\bibinfo {volume} {106}},\ \bibinfo {pages} {025018} (\bibinfo {year} {2022})}\BibitemShut {NoStop}%
\bibitem [{\citenamefont {Raychaudhuri}(1955)}]{Raychaudhuri:1955}%
  \BibitemOpen
  \bibfield  {author} {\bibinfo {author} {\bibfnamefont {A.}~\bibnamefont {Raychaudhuri}},\ }\bibfield  {title} {\bibinfo {title} {Relativistic cosmology. {I}},\ }\href {https://doi.org/10.1103/PhysRev.98.1123} {\bibfield  {journal} {\bibinfo  {journal} {Phys. Rev.}\ }\textbf {\bibinfo {volume} {98}},\ \bibinfo {pages} {1123} (\bibinfo {year} {1955})}\BibitemShut {NoStop}%
\bibitem [{\citenamefont {Hawking}\ and\ \citenamefont {Ellis}(1973)}]{Hawking:1973}%
  \BibitemOpen
  \bibfield  {author} {\bibinfo {author} {\bibfnamefont {S.~W.}\ \bibnamefont {Hawking}}\ and\ \bibinfo {author} {\bibfnamefont {G.~F.~R.}\ \bibnamefont {Ellis}},\ }\href {https://doi.org/10.1017/CBO9780511524646} {\emph {\bibinfo {title} {The Large Scale Structure of Space-Time}}}\ (\bibinfo  {publisher} {Cambridge University Press},\ \bibinfo {address} {Cambridge, United Kingdom},\ \bibinfo {year} {1973})\BibitemShut {NoStop}%
\bibitem [{\citenamefont {Garc\'{\i}a-Moreno}\ and\ \citenamefont {Jim\'enez~Cano}(2024)}]{Garcia:2023}%
  \BibitemOpen
  \bibfield  {author} {\bibinfo {author} {\bibfnamefont {G.}~\bibnamefont {Garc\'{\i}a-Moreno}}\ and\ \bibinfo {author} {\bibfnamefont {A.}~\bibnamefont {Jim\'enez~Cano}},\ }\bibfield  {title} {\bibinfo {title} {Nonexistence of a parent theory for general relativity and unimodular gravity},\ }\href {https://doi.org/10.1103/PhysRevD.109.104004} {\bibfield  {journal} {\bibinfo  {journal} {Phys. Rev. D}\ }\textbf {\bibinfo {volume} {109}},\ \bibinfo {pages} {104004} (\bibinfo {year} {2024})}\BibitemShut {NoStop}%
\bibitem [{\citenamefont {Montesinos}\ and\ \citenamefont {Gonzalez}(2023)}]{Montesinos:2023}%
  \BibitemOpen
  \bibfield  {author} {\bibinfo {author} {\bibfnamefont {M.}~\bibnamefont {Montesinos}}\ and\ \bibinfo {author} {\bibfnamefont {D.}~\bibnamefont {Gonzalez}},\ }\bibfield  {title} {\bibinfo {title} {Diffeomorphism-invariant action principles for trace-free {E}instein gravity},\ }\href {https://doi.org/10.1103/PhysRevD.108.124013} {\bibfield  {journal} {\bibinfo  {journal} {Phys. Rev. D}\ }\textbf {\bibinfo {volume} {108}},\ \bibinfo {pages} {124013} (\bibinfo {year} {2023})}\BibitemShut {NoStop}%
\bibitem [{\citenamefont {Lashkari}\ \emph {et~al.}(2014)\citenamefont {Lashkari}, \citenamefont {McDermott},\ and\ \citenamefont {Van~Raamsdonk}}]{Lashkari:2014}%
  \BibitemOpen
  \bibfield  {author} {\bibinfo {author} {\bibfnamefont {N.}~\bibnamefont {Lashkari}}, \bibinfo {author} {\bibfnamefont {M.~B.}\ \bibnamefont {McDermott}},\ and\ \bibinfo {author} {\bibfnamefont {M.}~\bibnamefont {Van~Raamsdonk}},\ }\bibfield  {title} {\bibinfo {title} {Gravitational dynamics from entanglement "thermodynamics"},\ }\href {https://doi.org/10.1007/JHEP04(2014)195} {\bibfield  {journal} {\bibinfo  {journal} {J. High Energ. Phys.}\ }\textbf {\bibinfo {volume} {2014}}\bibinfo  {number} { (195)}}\BibitemShut {NoStop}%
\bibitem [{\citenamefont {Faulkner}\ \emph {et~al.}(2014)\citenamefont {Faulkner}, \citenamefont {Guica}, \citenamefont {Hartman}, \citenamefont {Myers},\ and\ \citenamefont {Van~Raamsdonk}}]{Faulkner:2014}%
  \BibitemOpen
\bibfield  {number} {  }\bibfield  {author} {\bibinfo {author} {\bibfnamefont {T.}~\bibnamefont {Faulkner}}, \bibinfo {author} {\bibfnamefont {M.}~\bibnamefont {Guica}}, \bibinfo {author} {\bibfnamefont {T.}~\bibnamefont {Hartman}}, \bibinfo {author} {\bibfnamefont {R.~C.}\ \bibnamefont {Myers}},\ and\ \bibinfo {author} {\bibfnamefont {M.}~\bibnamefont {Van~Raamsdonk}},\ }\bibfield  {title} {\bibinfo {title} {Gravitation from entanglement in holographic {C}{F}{T}s},\ }\href {https://doi.org/10.1007/JHEP03(2014)051} {\bibfield  {journal} {\bibinfo  {journal} {J. High Energ. Phys.}\ }\textbf {\bibinfo {volume} {2014}}\bibinfo  {number} { (051)}}\BibitemShut {NoStop}%
\bibitem [{\citenamefont {Faulkner}\ \emph {et~al.}(2017)\citenamefont {Faulkner}, \citenamefont {Haehl}, \citenamefont {Hijano}, \citenamefont {Parrikar}, \citenamefont {Rabideau},\ and\ \citenamefont {Van~Raamsdonk}}]{Faulkner:2017}%
  \BibitemOpen
\bibfield  {number} {  }\bibfield  {author} {\bibinfo {author} {\bibfnamefont {T.}~\bibnamefont {Faulkner}}, \bibinfo {author} {\bibfnamefont {F.~M.}\ \bibnamefont {Haehl}}, \bibinfo {author} {\bibfnamefont {E.}~\bibnamefont {Hijano}}, \bibinfo {author} {\bibfnamefont {O.}~\bibnamefont {Parrikar}}, \bibinfo {author} {\bibfnamefont {C.}~\bibnamefont {Rabideau}},\ and\ \bibinfo {author} {\bibfnamefont {M.}~\bibnamefont {Van~Raamsdonk}},\ }\bibfield  {title} {\bibinfo {title} {Nonlinear gravity from entanglement in conformal field theories},\ }\href {https://doi.org/10.1007/JHEP08(2017)057} {\bibfield  {journal} {\bibinfo  {journal} {J. High Energ. Phys.}\ }\textbf {\bibinfo {volume} {2017}}\bibinfo  {number} { (057)}}\BibitemShut {NoStop}%
\bibitem [{\citenamefont {Ryu}\ and\ \citenamefont {Takayanagi}(2006)}]{Ryu:2006}%
  \BibitemOpen
\bibfield  {number} {  }\bibfield  {author} {\bibinfo {author} {\bibfnamefont {S.}~\bibnamefont {Ryu}}\ and\ \bibinfo {author} {\bibfnamefont {T.}~\bibnamefont {Takayanagi}},\ }\bibfield  {title} {\bibinfo {title} {Holographic derivation of entanglement entropy from the anti--de {S}itter space/conformal field theory correspondence},\ }\href {https://doi.org/10.1103/PhysRevLett.96.181602} {\bibfield  {journal} {\bibinfo  {journal} {Phys. Rev. Lett.}\ }\textbf {\bibinfo {volume} {96}},\ \bibinfo {pages} {181602} (\bibinfo {year} {2006})}\BibitemShut {NoStop}%
\bibitem [{\citenamefont {Casini}\ \emph {et~al.}(2016)\citenamefont {Casini}, \citenamefont {Galante},\ and\ \citenamefont {Myers}}]{Casini:2016}%
  \BibitemOpen
  \bibfield  {author} {\bibinfo {author} {\bibfnamefont {H.}~\bibnamefont {Casini}}, \bibinfo {author} {\bibfnamefont {D.~A.}\ \bibnamefont {Galante}},\ and\ \bibinfo {author} {\bibfnamefont {R.~C.}\ \bibnamefont {Myers}},\ }\bibfield  {title} {\bibinfo {title} {Comments on {J}acobson’s ``{E}ntanglement equilibrium and the {E}instein equation''},\ }\href {https://doi.org/10.1007/JHEP03(2016)194} {\bibfield  {journal} {\bibinfo  {journal} {J. High Energ. Phys.}\ }\textbf {\bibinfo {volume} {2016}}\bibinfo  {number} { (194)}}\BibitemShut {NoStop}%
\bibitem [{\citenamefont {Speranza}(2016)}]{Speranza:2016}%
  \BibitemOpen
\bibfield  {number} {  }\bibfield  {author} {\bibinfo {author} {\bibfnamefont {A.~J.}\ \bibnamefont {Speranza}},\ }\bibfield  {title} {\bibinfo {title} {Entanglement entropy of excited states in conformal perturbation theory and the {E}instein equation},\ }\href {https://doi.org/10.1007/JHEP04(2016)105} {\bibfield  {journal} {\bibinfo  {journal} {J. High Energ. Phys.}\ }\textbf {\bibinfo {volume} {2016}}\bibinfo  {number} { (105)}}\BibitemShut {NoStop}%
\bibitem [{\citenamefont {Arias}\ \emph {et~al.}(2017)\citenamefont {Arias}, \citenamefont {Blanco}, \citenamefont {Casini},\ and\ \citenamefont {Huerta}}]{Arias:2016}%
  \BibitemOpen
\bibfield  {number} {  }\bibfield  {author} {\bibinfo {author} {\bibfnamefont {R.~E.}\ \bibnamefont {Arias}}, \bibinfo {author} {\bibfnamefont {D.~D.}\ \bibnamefont {Blanco}}, \bibinfo {author} {\bibfnamefont {H.}~\bibnamefont {Casini}},\ and\ \bibinfo {author} {\bibfnamefont {M.}~\bibnamefont {Huerta}},\ }\bibfield  {title} {\bibinfo {title} {Local temperatures and local terms in modular {H}amiltonians},\ }\href {https://doi.org/10.1103/PhysRevD.95.065005} {\bibfield  {journal} {\bibinfo  {journal} {Phys. Rev. D}\ }\textbf {\bibinfo {volume} {95}},\ \bibinfo {pages} {065005} (\bibinfo {year} {2017})}\BibitemShut {NoStop}%
\bibitem [{\citenamefont {Arzano}(2020)}]{Arzano:2020}%
  \BibitemOpen
  \bibfield  {author} {\bibinfo {author} {\bibfnamefont {M.}~\bibnamefont {Arzano}},\ }\bibfield  {title} {\bibinfo {title} {Conformal quantum mechanics of causal diamonds},\ }\href {https://doi.org/10.1007/JHEP05(2020)072} {\bibfield  {journal} {\bibinfo  {journal} {J. High Energ. Phys.}\ }\textbf {\bibinfo {volume} {2020}}\bibinfo  {number} { (72)}}\BibitemShut {NoStop}%
\bibitem [{\citenamefont {Perez}\ and\ \citenamefont {Ribisi}(2024)}]{Ribisi:2023}%
  \BibitemOpen
\bibfield  {number} {  }\bibfield  {author} {\bibinfo {author} {\bibfnamefont {A.}~\bibnamefont {Perez}}\ and\ \bibinfo {author} {\bibfnamefont {S.}~\bibnamefont {Ribisi}},\ }\bibfield  {title} {\bibinfo {title} {Light-cone thermodynamics: Purification of the {M}inkowski vacuum},\ }\href {https://doi.org/10.1103/PhysRevD.109.125012} {\bibfield  {journal} {\bibinfo  {journal} {Phys. Rev. D}\ }\textbf {\bibinfo {volume} {109}},\ \bibinfo {pages} {125012} (\bibinfo {year} {2024})}\BibitemShut {NoStop}%
\bibitem [{\citenamefont {Chakraborty}\ \emph {et~al.}()\citenamefont {Chakraborty}, \citenamefont {Ord\'{o}\~{n}ez},\ and\ \citenamefont {Valdivia-Mera}}]{Chakraborty:2023}%
  \BibitemOpen
  \bibfield  {author} {\bibinfo {author} {\bibfnamefont {A.}~\bibnamefont {Chakraborty}}, \bibinfo {author} {\bibfnamefont {C.~R.}\ \bibnamefont {Ord\'{o}\~{n}ez}},\ and\ \bibinfo {author} {\bibfnamefont {G.}~\bibnamefont {Valdivia-Mera}},\ }\href {https://doi.org/10.48550/arXiv.2312.03541} {\bibinfo {title} {Path integral derivation of the thermofield double state in causal diamonds}},\ \Eprint {https://arxiv.org/abs/2312.03541} {arxiv:2312.03541 [hep-th]} \BibitemShut {NoStop}%
\bibitem [{\citenamefont {Jacobson}\ \emph {et~al.}(2017)\citenamefont {Jacobson}, \citenamefont {Senovilla},\ and\ \citenamefont {Speranza}}]{Jacobson:2017}%
  \BibitemOpen
  \bibfield  {author} {\bibinfo {author} {\bibfnamefont {T.}~\bibnamefont {Jacobson}}, \bibinfo {author} {\bibfnamefont {J.~M.~M.}\ \bibnamefont {Senovilla}},\ and\ \bibinfo {author} {\bibfnamefont {A.~J.}\ \bibnamefont {Speranza}},\ }\bibfield  {title} {\bibinfo {title} {Area deficits and the {B}el-{R}obinson tensor},\ }\bibfield  {journal} {\bibinfo  {journal} {Class. Quant. Grav.}\ }\textbf {\bibinfo {volume} {35}},\ \href {https://doi.org/10.1088/1361-6382/aab06e} {10.1088/1361-6382/aab06e} (\bibinfo {year} {2017})\BibitemShut {NoStop}%
\bibitem [{\citenamefont {Brewin}(2009)}]{Brewin:2009}%
  \BibitemOpen
  \bibfield  {author} {\bibinfo {author} {\bibfnamefont {L.}~\bibnamefont {Brewin}},\ }\bibfield  {title} {\bibinfo {title} {Riemann normal coordinate expansions using {C}adabra},\ }\href {https://doi.org/10.1088/0264-9381/26/17/175017} {\bibfield  {journal} {\bibinfo  {journal} {Class. Quant. Grav.}\ }\textbf {\bibinfo {volume} {26}},\ \bibinfo {pages} {175017} (\bibinfo {year} {2009})}\BibitemShut {NoStop}%
\bibitem [{\citenamefont {Lee}\ and\ \citenamefont {Wald}(1990)}]{Wald:1990}%
  \BibitemOpen
  \bibfield  {author} {\bibinfo {author} {\bibfnamefont {J.}~\bibnamefont {Lee}}\ and\ \bibinfo {author} {\bibfnamefont {R.~M.}\ \bibnamefont {Wald}},\ }\bibfield  {title} {\bibinfo {title} {Local symmetries and constraints},\ }\href {https://doi.org/10.1063/1.528801} {\bibfield  {journal} {\bibinfo  {journal} {J. Math. Phys.}\ }\textbf {\bibinfo {volume} {31}},\ \bibinfo {pages} {725} (\bibinfo {year} {1990})}\BibitemShut {NoStop}%
\bibitem [{\citenamefont {Bousso}\ \emph {et~al.}(2016)\citenamefont {Bousso}, \citenamefont {Fisher}, \citenamefont {Leichenauer},\ and\ \citenamefont {Wall}}]{Bousso:2016}%
  \BibitemOpen
  \bibfield  {author} {\bibinfo {author} {\bibfnamefont {R.}~\bibnamefont {Bousso}}, \bibinfo {author} {\bibfnamefont {Z.}~\bibnamefont {Fisher}}, \bibinfo {author} {\bibfnamefont {S.}~\bibnamefont {Leichenauer}},\ and\ \bibinfo {author} {\bibfnamefont {A.~C.}\ \bibnamefont {Wall}},\ }\bibfield  {title} {\bibinfo {title} {Quantum focusing conjecture},\ }\href {https://doi.org/10.1103/PhysRevD.93.064044} {\bibfield  {journal} {\bibinfo  {journal} {Phys. Rev. D}\ }\textbf {\bibinfo {volume} {93}},\ \bibinfo {pages} {064044} (\bibinfo {year} {2016})}\BibitemShut {NoStop}%
\bibitem [{\citenamefont {Jacobson}\ and\ \citenamefont {Visser}(2019{\natexlab{b}})}]{Jacobson:2019b}%
  \BibitemOpen
  \bibfield  {author} {\bibinfo {author} {\bibfnamefont {T.}~\bibnamefont {Jacobson}}\ and\ \bibinfo {author} {\bibfnamefont {M.~R.}\ \bibnamefont {Visser}},\ }\bibfield  {title} {\bibinfo {title} {Spacetime equilibrium at negative temperature and the attraction of gravity},\ }\href {https://doi.org/10.1142/S0218271819440164} {\bibfield  {journal} {\bibinfo  {journal} {Int. J. Mod. Phys. D}\ }\textbf {\bibinfo {volume} {28}},\ \bibinfo {pages} {194016} (\bibinfo {year} {2019}{\natexlab{b}})}\BibitemShut {NoStop}%
\bibitem [{\citenamefont {Tavlayan}\ and\ \citenamefont {Tekin}(2023)}]{Tavlayan:2023}%
  \BibitemOpen
  \bibfield  {author} {\bibinfo {author} {\bibfnamefont {A.}~\bibnamefont {Tavlayan}}\ and\ \bibinfo {author} {\bibfnamefont {B.}~\bibnamefont {Tekin}},\ }\bibfield  {title} {\bibinfo {title} {Partition function of a volume of space in a higher curvature theory},\ }\href {https://doi.org/10.1103/PhysRevD.108.L041902} {\bibfield  {journal} {\bibinfo  {journal} {Phys. Rev. D}\ }\textbf {\bibinfo {volume} {108}},\ \bibinfo {pages} {L041902} (\bibinfo {year} {2023})}\BibitemShut {NoStop}%
\bibitem [{\citenamefont {Unruh}(1989)}]{Unruh:1989}%
  \BibitemOpen
  \bibfield  {author} {\bibinfo {author} {\bibfnamefont {W.~G.}\ \bibnamefont {Unruh}},\ }\bibfield  {title} {\bibinfo {title} {Unimodular theory of canonical quantum gravity},\ }\href {https://doi.org/10.1103/PhysRevD.40.1048} {\bibfield  {journal} {\bibinfo  {journal} {Phys. Rev. D}\ }\textbf {\bibinfo {volume} {40}},\ \bibinfo {pages} {1048} (\bibinfo {year} {1989})}\BibitemShut {NoStop}%
\bibitem [{\citenamefont {Finkelstein}\ \emph {et~al.}(2001)\citenamefont {Finkelstein}, \citenamefont {Galiautdinov},\ and\ \citenamefont {Baugh}}]{Finkelstein:2001}%
  \BibitemOpen
  \bibfield  {author} {\bibinfo {author} {\bibfnamefont {D.~R.}\ \bibnamefont {Finkelstein}}, \bibinfo {author} {\bibfnamefont {A.~A.}\ \bibnamefont {Galiautdinov}},\ and\ \bibinfo {author} {\bibfnamefont {J.~E.}\ \bibnamefont {Baugh}},\ }\bibfield  {title} {\bibinfo {title} {Unimodular relativity and cosmological constant},\ }\href {https://doi.org/10.1063/1.1328077} {\bibfield  {journal} {\bibinfo  {journal} {J. Math. Phys.}\ }\textbf {\bibinfo {volume} {42}},\ \bibinfo {pages} {340} (\bibinfo {year} {2001})}\BibitemShut {NoStop}%
\bibitem [{\citenamefont {Reeh}\ and\ \citenamefont {Schlieder}(1961)}]{Reeh:1961}%
  \BibitemOpen
  \bibfield  {author} {\bibinfo {author} {\bibfnamefont {H.}~\bibnamefont {Reeh}}\ and\ \bibinfo {author} {\bibfnamefont {S.}~\bibnamefont {Schlieder}},\ }\bibfield  {title} {\bibinfo {title} {Bemerkungen zur unitäräquivalenz von lorentzinvarianten feldern.},\ }\href {https://doi.org/10.1007/BF02787889} {\bibfield  {journal} {\bibinfo  {journal} {Nuov. Cim.}\ }\textbf {\bibinfo {volume} {22}},\ \bibinfo {pages} {1051} (\bibinfo {year} {1961})}\BibitemShut {NoStop}%
\bibitem [{\citenamefont {Agullo}\ \emph {et~al.}(2023)\citenamefont {Agullo}, \citenamefont {Bonga}, \citenamefont {Ribes-Metidieri}, \citenamefont {Kranas},\ and\ \citenamefont {Nadal-Gisbert}}]{Agullo:2023}%
  \BibitemOpen
  \bibfield  {author} {\bibinfo {author} {\bibfnamefont {I.}~\bibnamefont {Agullo}}, \bibinfo {author} {\bibfnamefont {B.}~\bibnamefont {Bonga}}, \bibinfo {author} {\bibfnamefont {P.}~\bibnamefont {Ribes-Metidieri}}, \bibinfo {author} {\bibfnamefont {D.}~\bibnamefont {Kranas}},\ and\ \bibinfo {author} {\bibfnamefont {S.}~\bibnamefont {Nadal-Gisbert}},\ }\bibfield  {title} {\bibinfo {title} {How ubiquitous is entanglement in quantum field theory?},\ }\href {https://doi.org/10.1103/PhysRevD.108.085005} {\bibfield  {journal} {\bibinfo  {journal} {Phys. Rev. D}\ }\textbf {\bibinfo {volume} {108}},\ \bibinfo {pages} {085005} (\bibinfo {year} {2023})}\BibitemShut {NoStop}%
\bibitem [{\citenamefont {Susskind}\ and\ \citenamefont {Uglum}(1994)}]{Susskind:1994}%
  \BibitemOpen
  \bibfield  {author} {\bibinfo {author} {\bibfnamefont {L.}~\bibnamefont {Susskind}}\ and\ \bibinfo {author} {\bibfnamefont {J.}~\bibnamefont {Uglum}},\ }\bibfield  {title} {\bibinfo {title} {Black hole entropy in canonical quantum gravity and superstring theory},\ }\href {https://doi.org/10.1103/PhysRevD.50.2700} {\bibfield  {journal} {\bibinfo  {journal} {Phys. Rev. D}\ }\textbf {\bibinfo {volume} {50}},\ \bibinfo {pages} {2700} (\bibinfo {year} {1994})}\BibitemShut {NoStop}%
\bibitem [{\citenamefont {Jacobson}()}]{Jacobson:1994}%
  \BibitemOpen
  \bibfield  {author} {\bibinfo {author} {\bibfnamefont {T.}~\bibnamefont {Jacobson}},\ }\href {https://doi.org/10.48550/arXiv.gr-qc/9404039} {\bibinfo {title} {Black hole entropy and induced gravity}},\ \Eprint {https://arxiv.org/abs/2401.03572} {arxiv:2401.03572 [gr-qc]} \BibitemShut {NoStop}%
\bibitem [{\citenamefont {'t~Hooft}(1985)}]{Hooft:1985}%
  \BibitemOpen
  \bibfield  {author} {\bibinfo {author} {\bibfnamefont {G.}~\bibnamefont {'t~Hooft}},\ }\bibfield  {title} {\bibinfo {title} {On the quantum structure of a black hole},\ }\href {https://doi.org/10.1016/0550-3213(85)90418-3} {\bibfield  {journal} {\bibinfo  {journal} {Nucl. Phys. B}\ }\textbf {\bibinfo {volume} {256}},\ \bibinfo {pages} {727} (\bibinfo {year} {1985})}\BibitemShut {NoStop}%
\bibitem [{\citenamefont {Demers}\ \emph {et~al.}(1995)\citenamefont {Demers}, \citenamefont {Lafrance},\ and\ \citenamefont {Myers}}]{Demers:1995}%
  \BibitemOpen
  \bibfield  {author} {\bibinfo {author} {\bibfnamefont {J.-G.}\ \bibnamefont {Demers}}, \bibinfo {author} {\bibfnamefont {R.}~\bibnamefont {Lafrance}},\ and\ \bibinfo {author} {\bibfnamefont {R.~C.}\ \bibnamefont {Myers}},\ }\bibfield  {title} {\bibinfo {title} {Black hole entropy without brick walls},\ }\href {https://doi.org/10.1103/PhysRevD.52.2245} {\bibfield  {journal} {\bibinfo  {journal} {Phys. Rev. D}\ }\textbf {\bibinfo {volume} {52}},\ \bibinfo {pages} {2245} (\bibinfo {year} {1995})}\BibitemShut {NoStop}%
\bibitem [{\citenamefont {Banks}\ \emph {et~al.}()\citenamefont {Banks}, \citenamefont {Draper},\ and\ \citenamefont {Karydas}}]{Banks:2024}%
  \BibitemOpen
  \bibfield  {author} {\bibinfo {author} {\bibfnamefont {T.}~\bibnamefont {Banks}}, \bibinfo {author} {\bibfnamefont {P.}~\bibnamefont {Draper}},\ and\ \bibinfo {author} {\bibfnamefont {M.}~\bibnamefont {Karydas}},\ }\href {https://doi.org/10.48550/arXiv.2401.03572} {\bibinfo {title} {Breakdown of field theory in near-horizon regions}},\ \Eprint {https://arxiv.org/abs/2401.03572} {arxiv:2401.03572 [hep-th]} \BibitemShut {NoStop}%
\bibitem [{\citenamefont {Dong}(2014)}]{Dong:2014}%
  \BibitemOpen
  \bibfield  {author} {\bibinfo {author} {\bibfnamefont {X.}~\bibnamefont {Dong}},\ }\bibfield  {title} {\bibinfo {title} {Holographic entanglement entropy for general higher derivative gravity},\ }\href {https://doi.org/10.1007/JHEP01(2014)044} {\bibfield  {journal} {\bibinfo  {journal} {J. High Energ. Phys.}\ }\textbf {\bibinfo {volume} {2014}}\bibinfo  {number} { (44)}}\BibitemShut {NoStop}%
\bibitem [{\citenamefont {Wall}(2015)}]{Wall:2015}%
  \BibitemOpen
\bibfield  {number} {  }\bibfield  {author} {\bibinfo {author} {\bibfnamefont {A.~C.}\ \bibnamefont {Wall}},\ }\bibfield  {title} {\bibinfo {title} {A second law for higher curvature gravity},\ }\href {https://doi.org/10.1142/S0218271815440149} {\bibfield  {journal} {\bibinfo  {journal} {Int. J. Mod. Phys. D}\ }\textbf {\bibinfo {volume} {24}},\ \bibinfo {pages} {1544014} (\bibinfo {year} {2015})}\BibitemShut {NoStop}%
\bibitem [{\citenamefont {Cardy}(1986)}]{Cardy:1986}%
  \BibitemOpen
  \bibfield  {author} {\bibinfo {author} {\bibfnamefont {J.~L.}\ \bibnamefont {Cardy}},\ }\bibfield  {title} {\bibinfo {title} {Operator content of two-dimensional conformally invariant theories},\ }\href {https://doi.org/10.1016/0550-3213(86)90552-3} {\bibfield  {journal} {\bibinfo  {journal} {Nucl. Phys. B}\ }\textbf {\bibinfo {volume} {270}},\ \bibinfo {pages} {186} (\bibinfo {year} {1986})}\BibitemShut {NoStop}%
\bibitem [{\citenamefont {Carlip}(1999)}]{Carlip:1999}%
  \BibitemOpen
  \bibfield  {author} {\bibinfo {author} {\bibfnamefont {S.}~\bibnamefont {Carlip}},\ }\bibfield  {title} {\bibinfo {title} {Black hole entropy from conformal field theory in any dimension},\ }\href {https://doi.org/10.1103/PhysRevLett.82.2828} {\bibfield  {journal} {\bibinfo  {journal} {Phys. Rev. Lett.}\ }\textbf {\bibinfo {volume} {82}},\ \bibinfo {pages} {2828} (\bibinfo {year} {1999})}\BibitemShut {NoStop}%
\end{thebibliography}%

\end{document}